\documentclass[twocolumn,aps]{revtex4-1}

\usepackage{graphicx}
\begin{document}

\title{Non-classicality criteria for N-dimensional optical fields detected by quadratic detectors}

\author{Jan Pe\v{r}ina Jr.}
\email{jan.perina.jr@upol.cz}
\affiliation{Institute of Physics of the Czech
Academy of Sciences, Joint Laboratory of Optics of Palack\'{y} University and
Institute of Physics of CAS, 17. listopadu 50a, 772 07 Olomouc, Czech Republic}

\author{V\' aclav Mich\' alek}
\affiliation{Joint Laboratory of Optics of Palack\'{y} University and Institute
of Physics of the Czech Academy of Sciences, Faculty of Science, Palack\'{y}
University, 17. listopadu 12, 77146 Olomouc, Czech Republic}

\author{Pavel Pavl\'\i\v{c}ek}
\affiliation{Institute of Physics of the Czech Academy of Sciences, Joint
Laboratory of Optics of Palack\'{y} University and Institute of Physics of CAS,
17. listopadu 50a, 772 07 Olomouc, Czech Republic}

\author{Radek Machulka}
\affiliation{Joint Laboratory of Optics of Palack\'{y} University and Institute
of Physics of the Czech Academy of Sciences, Faculty of Science, Palack\'{y}
University, 17. listopadu 12, 77146 Olomouc, Czech Republic}

\author{Ond\v{r}ej Haderka}
\affiliation{Joint Laboratory of Optics of Palack\'{y} University and Institute of Physics of the Czech Academy of
Sciences, Faculty of Science, Palack\'{y} University, 17. listopadu 12, 77146 Olomouc, Czech Republic}

\begin{abstract}
Non-classicality criteria for general $ N $-dimensional optical fields are
derived. They involve intensity moments, the probabilities of photon-number
distributions or combinations of both. The Hillery criteria for the sums of the
probabilities of even or odd photon numbers are generalized to $ N
$-dimensional fields. As an example, the derived non-classicality criteria are
applied to an experimental 3-mode optical field containing two types of
photon-pair contributions. The accompanying non-classicality depths are used to
mutually compare their performance.
\end{abstract}

\pacs{} \maketitle

\section{Introduction}

Identification of non-classicality of optical fields
\cite{Mandel1995} has a long-lasting tradition in quantum optics.
For many years, the non-classicality was identified and quantified
by specific physical quantities suitable for revealing the
non-classicality of different kinds of quantum fields
\cite{Mandel1995}. The Fano factor that quantifies the strength of
photon-number fluctuations and the principal squeeze variance
\cite{Luks1988} that measures the squeezing of field phase
fluctuations serve as typical examples. With the fast development
of quantum optics, a large variety of nonclassical fields has been
suggested and experimentally realized
\cite{Dodonov2002,Lvovsky2009}. Many of them have no distinct
properties that would allow to tailor specific non-classicality
criteria (NCCa) for them. For this reason, the NCCa started to be
investigated from the general point of view.

Here, we address the NCCa based on the measurement of
photon-number distributions \cite{Perina1991} by quadratic
detectors. Though these criteria are not sensitive to the phases
of optical fields, they are extraordinarily useful for
(spatio-spectrally) multi-mode optical fields. The multi-mode
character of these fields implies that their statistical
properties are fully characterized by photon-number measurements.
The NCCa are traditionally written as the non-classicality
inequalities that involve the moments of integrated intensities.
These integrated-intensity moments denote the normally-ordered
photon number moments \cite{Perina1991} that are derived from the
photon-number distributions and their usual photon-number moments
with the help of the commutation relations \cite{PerinaJr2020a}. A
great deal of attention has been devoted to such NCCa for
1-dimensional (1D) \cite{Short1983,Teich1985,Lee1990a} and
2-dimensional (2D)
\cite{Allevi2012,Allevi2013,Sperling2015,Harder2016,MaganaLoaiza2019,PerinaJr2021,PerinaJr2021a}
optical fields for practical reasons. Among others, they allow to
identify the non-classicality of sub-Poissonian fields (1D) and
twin beams (2D). Useful NCCa were derived by several approaches
including the Cauchy-Schwarz inequality, majorization theory and
by using the classically nonnegative polynomials. They were
summarized in Ref.~\cite{Arkhipov2016c} for 1D and in
Ref.~\cite{PerinaJr2017a} for 2D optical fields.

Klyshko suggested in Ref.~\cite{Klyshko1996} the use of the Mandel detection
formula \cite{Mandel1995,Perina1991} in combination with the NCCa involving the
intensity moments to arrive at the NCCa written in the photon-number
probabilities. These criteria are in general more sensitive to the
non-classicality \cite{Waks2004,Waks2006,Wakui2014,PerinaJr2017c}. Some of them
also allow to identify the regions of photon-number distributions where the
non-classicality resides. They were extensively studied in
Ref.~\cite{PerinaJr2020} for 1D and Ref.~\cite{PerinaJr2020a} for 2D optical
fields.

The 1D and 2D NCCa represent special variants of the general NCCa
written for arbitrary dimensions and addressed in this paper. They
are applied to marginal 1D and 2D fields of the general $ N
$-dimensional fields obtained by tracing out over the remaining
dimensions. Thus the NCCa for dimensions greater than two are in
principle more general than those written in 1D and 2D. They are
indispensable for identification and quantification of the
non-classicality of fields exhibiting genuine higher-order quantum
correlations. This is topical for 3-dimensional (3D) fields
generated in the process of third-harmonic generation (triples of
photons) \cite{Chekhova2005}, cascaded second-order processes
\cite{Shalm2013,Hamel2014} or by post-selection
\cite{Alexander2020} (GHZ-like states).

Here, we provide the derivation of the NCCa for general $ N
$-dimensional optical fields by generalizing the approaches
applied earlier to 1D and 2D optical fields. We reveal general
relations among the derived NCCa. We show that the commonly used
NCCa originating in the matrix approach using $ 2\times 2 $
matrices form a subset inside the group of the NCCa stemming from
the Cauchy-Schwarz inequality. We further show that the NCCa
provided by the majorization theory can be decomposed into the
much simpler ones. These fundamental NCCa can easily be derived
assuming simple non-negative polynomials. In general, we identify
the fundamental non-classicality inequalities, i.e. the
inequalities that are not implied by other even simpler
non-classicality inequalities. We develop the approach for
deriving an $ N $-dimensional form of the Hillery non-classicality
criteria \cite{Hillery1985}. We also suggest a new type of the
NCCa - the hybrid NCCa. They contain the field intensity moments
in some dimensions while using the probabilities of photon-number
distributions in the remaining dimensions.

The NCCa allow not only to identify the non-classicality. They are
useful in quantifying the non-classicality when the Lee concept of
non-classicality depth (NCD) \cite{Lee1991} or non-classicality
counting parameter \cite{PerinaJr2019} are applied. This allows to
judge the performance of NCCa and identify the NCCa suitable for
revealing the non-classicality for given types of nonclassical
states. Moreover, when greater number of the NCCa is applied, the
maximal reached NCDs are assumed to give fair estimate of the
overall non-classicality of the analyzed state. This is important
in numerous applications that exploit nonclassical states.

We note that the NCCa can also be applied to optical fields that
are modified before detection. This improves the ability of the
NCCa to reveal the non-classicality. Mixing of an analyzed field
with a known coherent state at a beam-splitter serves as a typical
example \cite{Kuhn2017,Arkhipov2018b}.

To demonstrate the performance of the derived NCCa, we apply the
derived NCCa to a 3D optical field that is built from two
photon-pair contributions. This example allows us to demonstrate
the main features of the derived NCCa.

The paper is organized as follows. We bring the intensity NCCa in Sec.~II whose
subsections are devoted to specific kinds of the NCCa. The probability NCCa are
summarized in Sec.~III. Hybrid criteria are introduced in Sec.~IV. The general
derivation of the Hillery criteria is contained in Sec.~V. Section~VI is
devoted to non-classicality quantification. As an example, an experimental 3D
optical field with pairwise correlations is analyzed in Sec.~VII. Conclusions
are drawn in Sec.~VIII.

\section{Intensity non-classicality criteria}

In this section, we pay attention to the non-classicality
inequalities that use the (integrated) intensity moments
\cite{Perina1991}. In the following subsections, we apply the
Cauchy-Schwarz inequality, the matrix approach, the majorization
theory and the method that uses nonnegative polynomials to arrive
at the intensity NCCa in $ N $ dimensions. We consider $ N $
fields described by their intensities $ W_j $ and photon numbers $
n_j $, $ j=1,\ldots,N $. To formally simplify the description we
write the intensities $ W_j $ of the constituting fields as the
elements of an $ N $-dimensional intensity vector $ {\bf W} \equiv
\{W_1,\ldots,W_N\} $. Similarly the corresponding photon numbers $
n_j $ occurring in the detection of the constituting fields are
conveniently arranged into a photon-number vector $ {\bf n} \equiv
\{n_1,\ldots,n_N\} $. The joint state of these fields is
characterized by the joint quasi-distribution $ P({\bf W}) $ of
integrated intensities and joint photon-number distribution $
p({\bf n}) $. Intensity moments of the overall field can then be
expressed as $ \langle {\bf W}^{\bf i}\rangle \equiv \langle
W_1^{i_1}\cdots W_N^{i_N} \rangle $ using the integer vector $
{\bf i} \equiv \{i_1,\ldots,i_N\} $ containing the powers in
individual dimensions. We also introduce the notation for $ N
$-dimensional factorial $ {\bf i}! \equiv i_1!\ldots i_N! $.

\subsection{Intensity criteria based on the Cauchy-Schwarz inequality}

For 2D optical fields, the Cauchy-Schwarz inequality provides a group of
powerful criteria \cite{PerinaJr2017c,PerinaJr2020a}. When we consider the real
functions $ f({\bf W}) = {\bf W}^{{\bf l}/2 + {\bf k}} $ and $ g({\bf W}) =
{\bf W}^{{\bf l}/2 + {\bf k}'} $ for arbitrary integer vectors $ {\bf l} $, $
{\bf k} $ and $ {\bf k}' $, the Cauchy-Schwarz inequality $ |\int f({\bf W})
g({\bf W}) P({\bf W})d{\bf W}|^2 \le \int f^2({\bf W}) P({\bf W}) d{\bf W} \int
g^2({\bf W}) P({\bf W}) d{\bf W} $ written for a classical nonnegative
probability distribution $ P({\bf W}) $ takes the form:
\begin{equation}  
 \langle {\bf W}^{{\bf l} + {\bf k} + {\bf k}'} \rangle^2 \le
  \langle {\bf W}^{{\bf l} + 2{\bf k} } \rangle
  \langle {\bf W}^{{\bf l} + 2{\bf k}' } \rangle,
\label{1}
\end{equation}
where we use the notation $ \langle \ldots \rangle = \int d{\bf W}
P({\bf W}) \ldots $ for the mean value and $ d{\bf W} \equiv
dW_1\cdots dW_N $. Introducing the integer vectors $ {\bf n} =
{\bf l} + {\bf k} + {\bf k}' $ and $ {\bf m} = {\bf l} + 2{\bf k}
$ we derive the following NCCa from the classical inequality
(\ref{1}):
\begin{equation}  
 C_{\bf n}^{\bf m} = \langle {\bf W}^{\bf m} \rangle
  \langle {\bf W}^{2{\bf n} - {\bf m}} \rangle - \langle {\bf W}^{\bf n} \rangle^2 < 0.
\label{2}
\end{equation}

\subsection{Intensity criteria based on the matrix approach}

Nonnegative quadratic forms are conveniently written in the matrix
form. Their consideration leads to powerful and versatile NCCa for
1D \cite{Arkhipov2016c} and 2D
\cite{Agarwal1992,Shchukin2005,Vogel2008,Miranowicz2010} optical
fields. Considering determinant of the $ 2\times 2 $ matrix
\begin{eqnarray} 
 & {\rm det} \langle \left[ \begin{array}{cc} {\bf W}^{2{\bf k}} & {\bf W}^{{\bf k}+{\bf l}} \\
 {\bf W}^{{\bf l}+{\bf k}} & {\bf W}^{2{\bf l}} \end{array} \right] \rangle =
  \langle {\bf W}^{2{\bf k}} \rangle \langle {\bf W}^{2{\bf l}}\rangle  -
  \langle {\bf W}^{{\bf k}+{\bf l}}\rangle^2  &
\label{3}
\end{eqnarray}
describing the quadratic form $ ({\bf W}^{\bf k} + {\bf W}^{\bf l})^2 $ we
arrive at the NCCa that form a subset of the group $ C_{\bf n}^{\bf m} $ in
Eq.~(\ref{2}) [$ {\bf n} \leftarrow {\bf k}+{\bf l} $, $ {\bf m} \leftarrow
2{\bf l} $].

On the other hand, the quadratic form $ ({\bf W}^{\bf k} + {\bf W}^{\bf l} +
{\bf W}^{\bf m})^2 $ encoded into the $ 3\times 3 $ matrix
\begin{eqnarray} 
  & \left[ \begin{array}{ccc} {\bf W}^{2{\bf k}} & {\bf W}^{{\bf k}+{\bf l}} & {\bf W}^{{\bf k}+{\bf m}}  \\
  {\bf W}^{{\bf l}+{\bf k}} & {\bf W}^{2{\bf l}} & {\bf W}^{{\bf l}+{\bf m}} \\
  {\bf W}^{{\bf m}+{\bf k}} & {\bf W}^{{\bf m}+{\bf l}} & {\bf W}^{2{\bf m}}
 \end{array} \right] &
\label{4}
\end{eqnarray}
provides the powerful NCCa with the complex structure:
\begin{eqnarray} 
 & M_{\bf klm} = \bigl\{ \langle {\bf W}^{2{\bf k}} \rangle [ \langle {\bf W}^{2{\bf l}}\rangle \langle {\bf W}^{2{\bf m}} \rangle
  - \langle {\bf W}^{{\bf l}+{\bf m}} \rangle^2 ]  & \nonumber \\
 & + ({\bf klm}) \leftarrow ({\bf lmk}) + ({\bf klm}) \leftarrow ({\bf mkl}) \bigr\}  & \nonumber \\
 & + 2[\langle {\bf W}^{{\bf k}+{\bf l}}  \rangle \langle {\bf W}^{{\bf k}+{\bf m}} \rangle \langle {\bf W}^{{\bf l}+{\bf m}} \rangle
   - \langle {\bf W}^{2{\bf k}} \rangle \langle {\bf W}^{2{\bf l}}\rangle \langle {\bf W}^{2{\bf m}}\rangle] &
 \nonumber \\
 &  <0.  &
\label{5}
\end{eqnarray}
Symbol $ \leftarrow $ used in Eq.~(\ref{5}) stands for the terms derived from
the explicitly written ones by the indicated change of indices.

\subsection{Intensity criteria originating in majorization theory and
nonnegative polynomials}

In the majorization theory \cite{Marshall2010} many classical
inequalities for polynomials written in fixed numbers of
independent variables are derived \cite{Lee1990b}. The
inequalities describing 'the movement of one ball towards the
right' \cite{Marshall2010} represent the basic building blocks of
the remaining inequalities. Assuming the integer vectors $ {\bf i}
$ and $ {\bf i}' = \{i_1,\ldots,i_k+1,\ldots,i_l-1,\ldots,i_N \}
$, $ k<l $, such that the vector $ {\bf i}' $ majorizes the vector
$ {\bf i} $ ($ {\bf i}' \succ {\bf i} $, $ i_1-1 \ge i_2-1
\ge\ldots \ge i_{k-1}-1 \ge i_{k} \ge \ldots \ge i_l \ge i_{l+1}+1
\ge \ldots \ge i_N+1 $) they give rise to the following
non-classicality inequalities:
\begin{equation}  
 \sum_{{\cal P}{\bf i}'} \langle {\bf W}^{{\bf i}'} \rangle <
  \sum_{{\cal P}{\bf i}} \langle {\bf W}^{{\bf i}} \rangle .
\label{6}
\end{equation}
In Eq.~(\ref{6}), symbol $ \cal P{\bf i} $ stands for all permutations of the
elements of vector $ \bf i $.

The non-classicality inequalities (\ref{6}) can be conveniently decomposed into
the much simpler ones expressed as:
\begin{eqnarray}  
 &E_{\bf i}^{kl} = \langle {\bf W}^{\bf i} (W_k-W_l)^2\rangle =
  \langle {\bf W}_{kl}^{{\bf i}_{kl}^\perp} (W_k^{i_k+2}W_l^{i_l} & \nonumber \\
 &  - 2W_k^{i_k+1}W_l^{i_l+1} + W_k^{i_k}W_l^{i_l+2}) \rangle < 0;&
\label{7}
\end{eqnarray}
$ {\bf W}_{kl} = W_1\ldots W_{k-1}W_{k+1}\ldots W_{l-1}W_{l+1}\ldots W_N $ and
$ {\bf i}_{kl}^\perp = \{
i_1,\ldots,i_{k-1},i_{k+1},\ldots,i_{l-1},i_{l+1},\ldots,i_N \} $. The
inequality (\ref{6}) can be expressed in terms of the simpler inequalities $
E_{\bf i}^{kl} $ as follows ($ i_k \ge i_l $):
\begin{eqnarray}  
 & \sum_{k'=1}^N \sum_{l'=1}^{k'-1} \sum_{{\cal P} {\bf i}_{k'l'}^\perp}
  \Bigl[ E_{{\cal P}_k^{k'} {\cal P}_l^{l'} \{ \ldots, i_k-1,\ldots,i_l-1,\ldots \} }^{k'l'}
  \nonumber \\
 & + E_{{\cal P}_k^{k'} {\cal P}_l^{l'} \{ \ldots, i_k-2,\ldots,i_l,\ldots \} }^{k'l'}
  + E_{{\cal P}_k^{k'} {\cal P}_l^{l'} \{ \ldots, i_k-3,\ldots,i_l+1,\ldots \}
  }^{k'l'} & \nonumber \\
 & + \ldots + E_{{\cal P}_k^{k'} {\cal P}_l^{l'} \{ \ldots, i_l+1,\ldots,i_k-3,\ldots \} }^{k'l'}
  + E_{{\cal P}_k^{k'} {\cal P}_l^{l'} \{ \ldots, i_l,\ldots,i_k-2,\ldots
  \}}^{k'l'} & \nonumber \\
 & + E_{{\cal P}_k^{k'} {\cal P}_l^{l'} \{ \ldots, i_l-1,\ldots,i_k-1,\ldots \} }^{k'l'}
  \Bigr] < 0 &
\label{8}
\end{eqnarray}
and $ {\cal P}_k^{k'} \{ \ldots, i_k,\ldots,i_{k'},\ldots \} = \{ \ldots,
i_{k'},\ldots,i_k,\ldots \} $.

As a consequence, any non-classicality inequality stemming from the
majorization theory is implied by the non-classicality inequalities $ E_{\bf
i}^{kl} $ given in Eq.~(\ref{7}). They can simply be generalized into the NCCa
\begin{equation}  
 E_{\bf i}^{\bf I} = \langle {\bf W}^{\bf i} \prod_{k=1}^N \prod_{l=k+1}^N (W_k-W_l)^{2I_{kl}} \rangle <
 0
\label{9}
\end{equation}
in which the integers $ I_{kl} $ form an upper triangular matrix $ {\bf I} $.

Also other NCCa based on nonnegative polynomials can be constructed. They are
useful for the states with specific forms of correlations. As an example we
write the NCCa $ P^k $ and $ P^{kl} $,
\begin{eqnarray}  
 P^k &=& \langle \left( \sum_{k'=1}^N W_{k'} -2W_k \right)^2 \rangle ,
\label{10} \\
 P^{kl} &=& \langle \left( \sum_{k'=1}^N W_{k'} -2W_k \right)^2 \left( \sum_{l'=1}^N W_{l'} -2W_l \right)^2 \rangle ,
\label{11}
\end{eqnarray}
useful for the non-classicality analysis of the states in Sec.~VI.

\section{Probability non-classicality criteria}

The Mandel detection formula \cite{Mandel1995,Perina1991} written for an $ N
$-dimensional optical field as
\begin{equation}  
 p({\bf n}) = \frac{1}{{\bf n}!} \int d{\bf W} {\bf W}^{\bf n} \exp(-\Sigma {\bf W})
  P({\bf W}),
\label{12}
\end{equation}
$ \Sigma {\bf W} = \sum_{k=1}^N W_k $, allows to derive the probability NCCa
from the corresponding intensity NCCa according to the following mapping (for
details, see the next section):
\begin{equation} 
  {\bf W}^{\bf n} \leftarrow {\bf n}! p({\bf n}) / p({\bf 0});
\label{13}
\end{equation}
symbol $ {\bf 0} $ denotes the vector with all elements equal to
0.

Using the mapping (\ref{13}), the intensity NCCa $ C_{\bf n}^{\bf m} $ from
Eq.~(\ref{3}) derived from the Cauchy-Schwarz inequality give the following
probability NCCa:
\begin{equation}  
 \bar{C}_{\bf n}^{\bf m} = \frac{(2\bf{n}-{\bf m})!{\bf m}! }{ ({\bf n}!)^2 } p({\bf m})p(2{\bf n}- {\bf m}) - p^2({\bf n}) <0.
\label{14}
\end{equation}

Similarly, the intensity NCCa $ M_{\bf klm} $ in Eq.~(\ref{5}) obtained by the
matrix approach are transformed into the following probability NCCa:
\begin{eqnarray} 
 & \bar{M}_{\bf klm} = \Bigl\{ p(2{\bf k})\Bigl[p(2{\bf l})p(2{\bf m})- \frac{({\bf l}+{\bf m})!^2}{(2{\bf l})!(2{\bf m})!} p^2({\bf l}+{\bf m})\Bigr] & \nonumber \\
 & + ({\bf klm}) \leftarrow ({\bf mkl}) + ({\bf klm}) \leftarrow ({\bf lmk}) \bigr\} & \nonumber \\
 & + 2\Bigl[ \frac{ ({\bf k}+{\bf l})!({\bf k}+{\bf m})!({\bf l}+{\bf m})!}{(2{\bf k})!(2{\bf l})!(2{\bf m})!} p({\bf k}+{\bf l})p({\bf k}+{\bf m}) & \nonumber \\
 & \times p({\bf l}+{\bf m}) - p(2{\bf k})p(2{\bf l})p(2{\bf m})\bigr] <0.&
\label{15}
\end{eqnarray}

Finally, the NCCa $ E_{\bf i}^{kl} $ from Eq.~(\ref{7}) give rise to the
following probability NCCa:
\begin{eqnarray}   
 & \bar{E}_{\bf i}^{kl} = (i_k+2)(i_k+1)p(\ldots,i_k+2,\ldots,i_l,\ldots) &
  \nonumber \\
 & + (i_l+2)(i_l+1)p(\ldots,i_k,\ldots,i_l+2,\ldots) & \nonumber \\
 & - 2(i_k+1)(i_l+1)p(\ldots,i_k+1,\ldots,i_l+1,\ldots) < 0. & \nonumber \\
 & &
\label{16}
\end{eqnarray}

\section{Hybrid non-classicality criteria}

In general, we may map via the mapping in Eq.~(\ref{13}) the
intensity moments to the probabilities only in some dimensions.
This leads to hybrid NCCa that combine the intensity moments and
probabilities. This results in the use of 'mixed moments' that are
determined along the formula
\begin{eqnarray}   
 & \langle {\bf W}_W^{{\bf i}_W}\rangle_{p_p({\bf i}_p)} =
  \int d{\bf W}_W \, {\bf W}_W^{{\bf i}_W} \frac{1}{ {\bf i}_p! } \int d{\bf W}_p \, {\bf W}_p^{{\bf i}_p} &
  \nonumber \\
 & \mbox{} \times  \exp(-\Sigma {\bf W}_p ) P({\bf W}_W,{\bf W}_p) , &
\label{17}
\end{eqnarray}
in which we split the elements of integer vector $ {\bf i} $ describing the
intensity-moment powers into two groups. In Eq.~(\ref{17}), we denote the
corresponding integer vectors by $ {\bf i}_W $ and $ {\bf i}_p $ and the
corresponding intensities by $ {\bf W}_W $ and $ {\bf W}_p $.

To reveal the mapping between the intensity and probability NCCa,
let us consider the normalized distribution $ P'({\bf W}_W,{\bf
W}_p) = \exp(-\Sigma {\bf W}_p ) P({\bf W}_W,{\bf W}_p) $ $ / [
\int d{\bf W}'_W \int d{\bf W}'_p \, \exp(-\Sigma {\bf W}'_p )
P({\bf W}'_W,{\bf W}'_p)] $ that stays nonnegative provided that
the original distribution $ P({\bf W}_W,{\bf W}_p) $ is
nonnegative. Thus the moments of this distribution $ \langle
\langle {\bf W}_W^{{\bf i}_W} {\bf W}_p^{{\bf i}_p}\rangle \rangle
$,
\begin{eqnarray}   
 & \langle \langle {\bf W}_W^{{\bf i}_W} {\bf W}_p^{{\bf i}_p} \rangle \rangle =
  \int d{\bf W}_W \int d{\bf W}_p \, {\bf W}_W^{{\bf i}_W} {\bf W}_p^{{\bf i}_p} &
  \nonumber \\
 & \mbox{} \times  P'({\bf W}_W,{\bf W}_p) , &
\label{18}
\end{eqnarray}
when substituted into the non-classicality inequalities with intensity moments
in Sec.~II give rise to the NCCa. When we express these NCCa in terms of the
'mixed moments' defined in Eq.~(\ref{17}) we reveal the following mapping
\begin{equation}  
 {\bf W}_W^{{\bf i}_W} {\bf W}_p^{{\bf i}_p} \leftarrow {\bf W}_W^{{\bf i}_W}
  {\bf i}_p! p_p({\bf i}_p) / p_p({\bf 0})
\label{19}
\end{equation}
in which the marginal photon-number distribution $ p_p $ is defined in the
dimensions grouped into the integer vector $ {\bf i}_p $:
\begin{eqnarray}   
 & p_p({\bf i}_p) = \frac{1}{ {\bf i}_p! }
   \int d{\bf W}_p {\bf W}_p^{{\bf i}_p} \exp(-\Sigma {\bf W}_p ) &
  \nonumber \\
 & \mbox{} \times \int d{\bf W}_W  P({\bf W}_W,{\bf W}_p) .
\label{20}
\end{eqnarray}

As an example of mapping in Eq.~(\ref{19}) we write the hybrid NCCa derived
from the intensity NCCa in Eq.~(\ref{2}) that originate in the Cauchy-Schwarz
inequality:
\begin{eqnarray}  
 & \tilde{C}_{{\bf n}_W,{\bf n}_p}^{{\bf m}_W,{\bf m}_p} =
  \frac{(2\bf{n}_p-{\bf m}_p)!{\bf m}_p! }{ ({\bf n}_p!)^2 }
  \langle {\bf W}_W^{2{\bf n}_W - {\bf m}_W} \rangle_{p_p(2{\bf n}_p - {\bf
  m}_p)} & \nonumber \\
 & \mbox{} \times \langle {\bf W}_W^{{\bf m}_W} \rangle_{p_p({\bf m}_p)}
  - \langle {\bf W}_W^{{\bf n}_W} \rangle_{p_p({\bf n}_p)}^2 < 0; &
\label{21}
\end{eqnarray}
the hybrid moments are determined along Eq.~(\ref{17}).

\section{Generalized Hillery non-classicality criteria}

The Hillery NCCa were derived for the sums of probabilities of even and odd
photon numbers for 1D optical fields \cite{Hillery1985}. They are suitable,
e.g., for evidencing the non-classicality of single-mode squeezed-vacuum
states.

To reveal their $ N $-dimensional generalizations, we first consider the
following classical inequality
\begin{equation} 
 \int d{\bf W}\, {\rm ch}(W_+) \exp(-W_+) P({\bf W}) \ge 1/2
\label{22}
\end{equation}
valid for any classical nonnegative distribution $ P({\bf W}) $. In
Eq.~(\ref{22}), $ {\rm ch} $ stands for the hyperbolic cosine and $ W_+ =
\Sigma {\bf W} $. A suitable unitary transformation to new variables that
involves the variable $ W_+ $ and replacement of $ {\rm ch} $ function by the
defining exponentials immediately reveal the inequality~(\ref{22}). The Taylor
expansion of $ {\rm ch} $ function and use of multinomial expansion transform
the inequality~(\ref{22}) into the form:
\begin{equation} 
 \sum_{l=0}^{\infty} \sum_{ {\bf n}={\bf 0}, \sum {\bf n} = 2l }^{\infty}
  \frac{1}{ {\bf n}! } \int d{\bf W} \, {\bf W}^{\bf n} \exp(-W_+) P({\bf W}) \ge 1/2 .
\label{23}
\end{equation}
The comparison of integrals in Eq.~(\ref{23}) with the Mandel
detection formula (\ref{12}) results in the following generalized
Hillery NCC $ H_1 $ for the probabilities of even-summed photon
numbers:
\begin{equation} 
 H_1 = \sum_{l=0}^{\infty} \sum_{{\bf n}={\bf 0}, \sum {\bf n} = 2l}^{\infty}
  p({\bf n}) - 1/2 < 0.
\label{24}
\end{equation}

The second Hillery NCC is obtained starting from the classical inequality
\begin{equation} 
 \int d{\bf W}\, [ {\rm sh}(W_+) - {\rm ch}(W_+) +1 ] \exp(-W_+) P({\bf W}) \ge 0
\label{225}
\end{equation}
in which $ {\rm sh} $ denotes the hyperbolic sine. Similarly as above, the use
of the Taylor expansion for $ {\rm sh} $ and $ {\rm ch} $ functions,
multinomial expansion and the Mandel detection formula leaves us with the
following NCC:
\begin{equation} 
 \sum_{l=0}^{\infty} \sum_{{\bf n}={\bf 0}, \sum {\bf n} = 1+2l}^{\infty}
  p({\bf n}) - \sum_{l=0}^{\infty} \sum_{{\bf n}={\bf 0}, \sum {\bf n} = 2l}^{\infty}
  p({\bf n}) + p({\bf 0}) < 0 .
\label{26}
\end{equation}
The use of the normalization condition $ \sum_{{\bf n}={\bf
0}}^{\infty} p({\bf n}) = 1 $ in Eq.~(\ref{26}) leads to the
second generalized Hillery NCC  $ H_2 $ for the probabilities of
odd-summed photon numbers:
\begin{equation} 
 H_2 = \sum_{l=0}^{\infty} \sum_{{\bf n}={\bf 0}, \sum {\bf n} = 1+2l}^{\infty}
  p({\bf n}) - [1-p({\bf 0})]/2 < 0.
\label{27}
\end{equation}

\section{Quantification of non-classicality}

The above written NCCa can be used not only as non-classicality
identifiers. When we apply the concept of the Lee non-classicality
depth \cite{Lee1991}, they also quantitatively characterize the
non-classicality. This quantification is based upon the properties
of optical fields described in different field-operator orderings
\cite{Perina199}. Whereas the normally-ordered intensity moments
contain only the intrinsic noise of the field, the general $ s
$-ordered intensity moments involve an additional 'detection'
thermal noise with the mean photon number $ (1-s)/2 $ per one mode
\cite{Perina1991,Lee1991}. Such noise gradually decreases the
non-classicality as the ordering parameter $ s $ decreases. The
threshold value $ s_{\rm th} $ at which a given field loses its
non-classicality then determines the non-classicality depth (NCD)
$ \tau $ \cite{Lee1991}:
\begin{equation} 
 \tau = \frac{1-s_{\rm th} }{2} .
\label{28}
\end{equation}
We note that, alternatively, we may apply the non-classicality counting
parameter introduced in Ref.~\cite{PerinaJr2019}.

To arrive at the NCDs $ \tau $ of the above written NCCa, we have to transform
the 'mixed moments' $ \langle {\bf W}_W^{{\bf i}_W}\rangle_{p_p({\bf i}_p)} $
of Eq.~(\ref{17}) to their general $ s $-ordered form $ \langle {\bf W}_W^{{\bf
i}_W}\rangle_{p_p({\bf i}_p),s} $. Using the results of
Refs.~\cite{Perina1991,PerinaJr2020a} the corresponding transformation is
accomplished via the matrices $ S_W $ and $ S_p $ appropriate to the intensity
moments and probabilities, respectively,
\begin{eqnarray}  
 & \langle {\bf W}_W^{{\bf i}_W}\rangle_{p_p({\bf i}_p),s} =
  \sum_{{\bf i}'_{W} = {\bf 1} }^{{\bf i}_{W}} S_W(i_{1_W},i'_{1_W};s,M_{1_W}) \ldots & \nonumber \\
 & S_W(i_{N_W},i'_{N_W};s,M_{N_W})
  \sum_{{\bf i}'_{p}={\bf 0} }^{\infty} S_p(i_{1_p},i'_{1_p};s,M_{1_p}) & \nonumber \\
 & \ldots S_p(i_{N_p},i'_{N_p};s,M_{N_p})
  \langle {\bf W}_W^{{\bf i}'_W}\rangle_{p_p({\bf i}'_p)}, &
\label{29}
\end{eqnarray}
$ {\bf 1} \equiv \{1,1,\ldots\} $ and integer vectors $ {\bf M}_W \equiv
\{M_{1_W},\ldots,M_{N_W} \} $ and $ {\bf M}_p \equiv \{M_{1_p},\ldots,M_{N_p}
\} $ give the numbers of modes in all dimensions of the analyzed optical field.
In Eq.~(\ref{29}), the transformation matrices $ S_W $ and $ S_p $ are defined
as follows:
\begin{eqnarray} 
 S_W(i,i';s,M) &=& \left(\begin{array}{c} i\\ i' \end{array}\right)
  \frac{\Gamma(i+M)}{\Gamma(i'+M)} \left(\frac{1-s}{2}\right)^{i-i'},
\label{30} \\
 S_p(i,i';s,M) &=& \left(\frac{2}{3-s}\right)^M
 \left(\frac{1+s}{1-s}\right)^{i'}
  \left(\frac{1-s}{3-s}\right)^i  \nonumber \\
  & & \hspace{0mm} \times \sum_{l=0}^{i'} (-1)^{i'-l} \left(\begin{array}{c} i' \\ l \end{array}\right)
  \left(\begin{array}{c} i+l+M-1\\ i \end{array}\right) \nonumber \\
  & & \times \left(\frac{4}{(1+s)(3-s)} \right)^l .
\label{31}
\end{eqnarray}

\section{Application of the derived non-classicality criteria to a 3-dimensional
optical field}

\begin{figure} 
 \resizebox{.7\hsize}{!}{\includegraphics{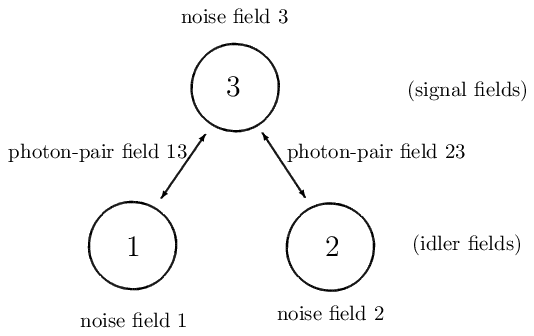}}
 \vspace{1mm}
 \caption{Structure of the analyzed 3D optical field composed of ideal photon-pair fields 13 and 23 and
  noise fields 1, 2 and 3.}
\label{fig1}
\end{figure}

\begin{figure} 
 \resizebox{.47\hsize}{!}{\includegraphics{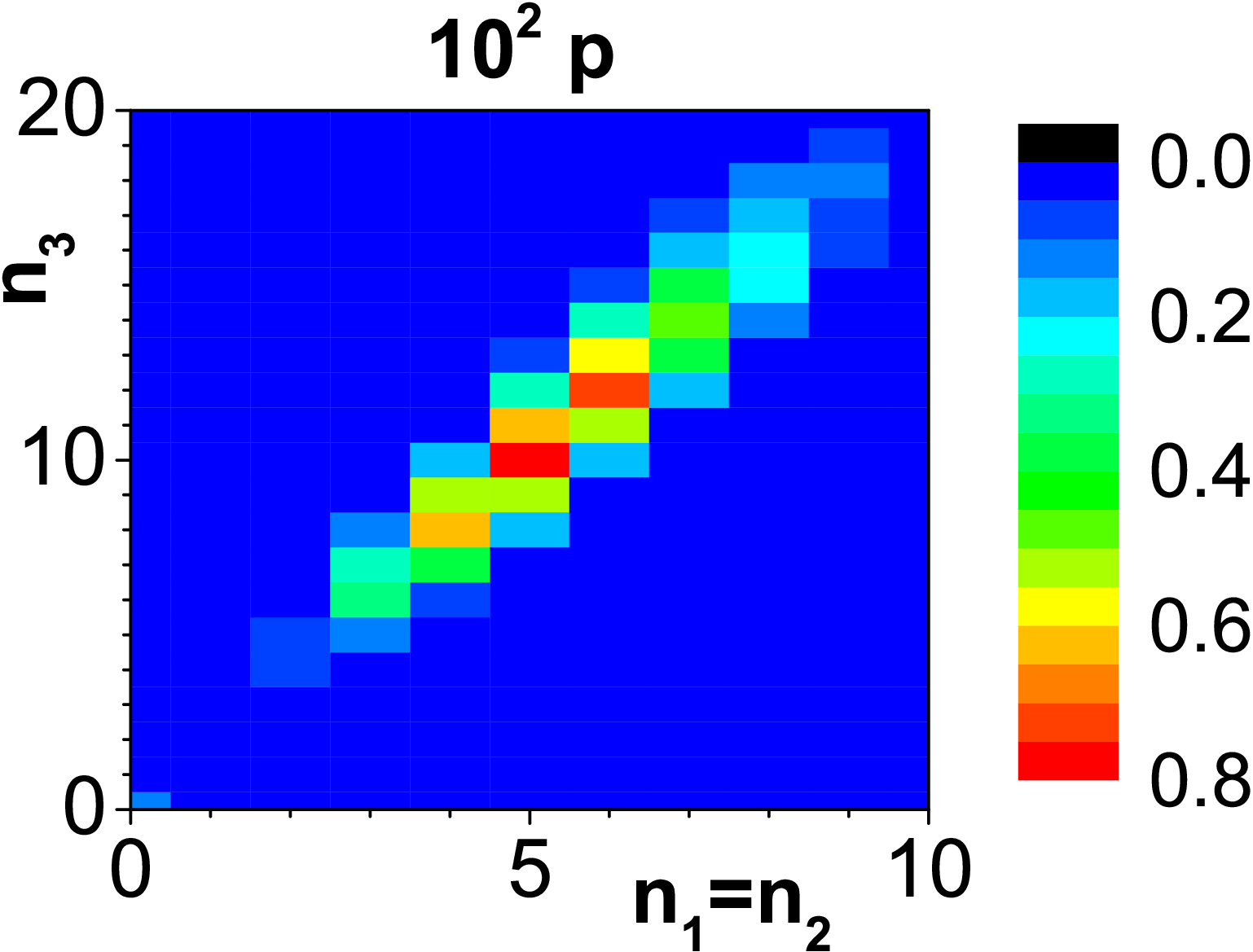}}  \hspace{1mm}
 \resizebox{.47\hsize}{!}{\includegraphics{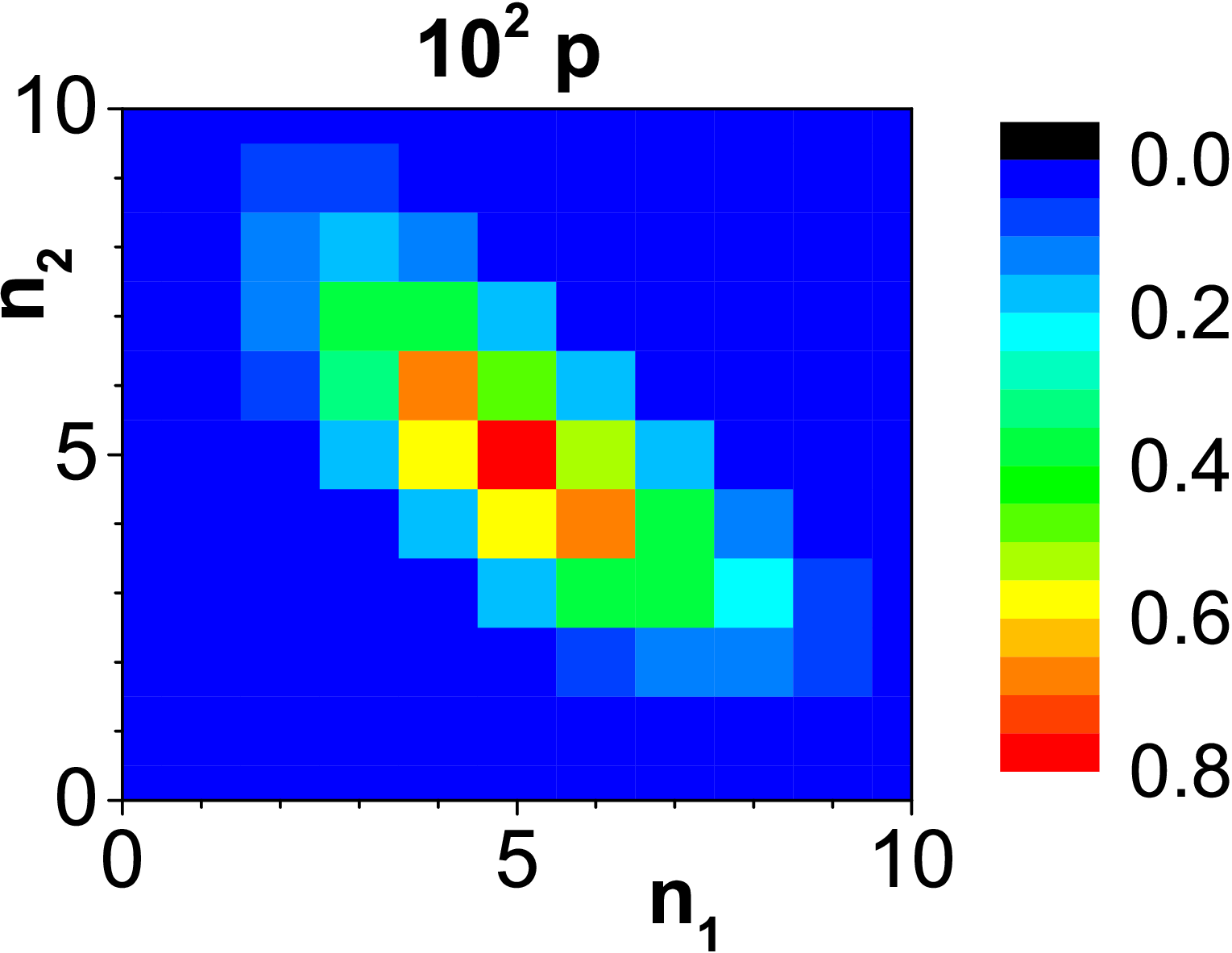}} \\
 \centerline{\small (a) \hspace{.4\hsize} (b)}
 \caption{Photon-number distribution $ p(n_1,n_2,n_3) $ obtained by ML approach in 2D cuts:
 (a) $ p(n_1=n_2,n_3) $ and (b) $ p(n_1,n_2,n_3=10) $. Relative errors are better than 1~\%.}
\label{fig2}
\end{figure}

We demonstrate the performance of the above written NCCa in
revealing the non-classicality using a 3D optical field containing
two types of photon pairs. This 3D field was generated in the
pulsed multimode spontaneous parametric down-conversion from two
nonlinear crystals that produced two types of photon pairs
differing in polarization (for details, see \cite{PerinaJr2021}).
While the idler fields constitute fields 1 and 2, the signal
fields overlap in a common detection area and together form field
3, as schematically shown in Fig.~\ref{fig1}.
The experimental photocount histogram $ f(c_1,c_2,c_3) $ that gives the
normalized number of simultaneous detections of $ c_i $ photocounts in field $
i $, $ i=1,2,3 $, in  $ 1.2 \times 10^{6} $ measurements \cite{PerinaJr2021}
was reconstructed using the maximum likelihood (ML) approach
\cite{Dempster1977,Vardi1993}. The iteration algorithm ($ j $ numbers the
iteration steps)
\begin{eqnarray}   
 p^{(j+1)}({\bf n})&=& \sum_{{\bf c}={\bf 0}}^{\infty} \frac{ f({\bf c}) T({\bf c},{\bf n}) }{
  \sum_{{\bf n}'={\bf 0}}^{\infty} T({\bf c},{\bf n}') p^{(j)}({\bf n}') }
\label{32}
\end{eqnarray}
provided the reconstructed 3D photon-number distribution $ p(n_1,n_2,n_3) $
that we analyze below from the point of view of its non-classicality. In
Eq.~(\ref{32}), the detection function $ T({\bf c},{\bf n}) $ gives the
probability of detecting $ {\bf c} $ photocounts at $ N $ detectors being
illuminated by $ {\bf n} $ photons. The theoretical prediction $ f^{\rm
th}({\bf c}) $ for the photocount histogram is obtained as $ f^{\rm th}({\bf
c}) = \sum_{{\bf n}={\bf 0}}^{\infty} T({\bf c},{\bf n}) p({\bf n}) $. The
analyzed photon-number distribution $ p $ is shown in Fig.~\ref{fig2} in two
characteristic cuts suitable for visualization of its pairwise structure.

The experimental 3D optical field was also fitted by the model of two ideal
multi-mode Gaussian twin beams ($ 6.15 \pm 0.05 $ and $ 5.95 \pm 0.05 $ mean
photon pairs) and three multi-mode Gaussian noise fields ($ 0.11\pm 0.02 $, $
0.07 \pm 0.01 $ and $ 0.02 \pm 0.01 $ mean noise photons). Details are found in
Ref.~\cite{PerinaJr2021}.

\begin{figure} 
 \resizebox{.47\hsize}{!}{\includegraphics{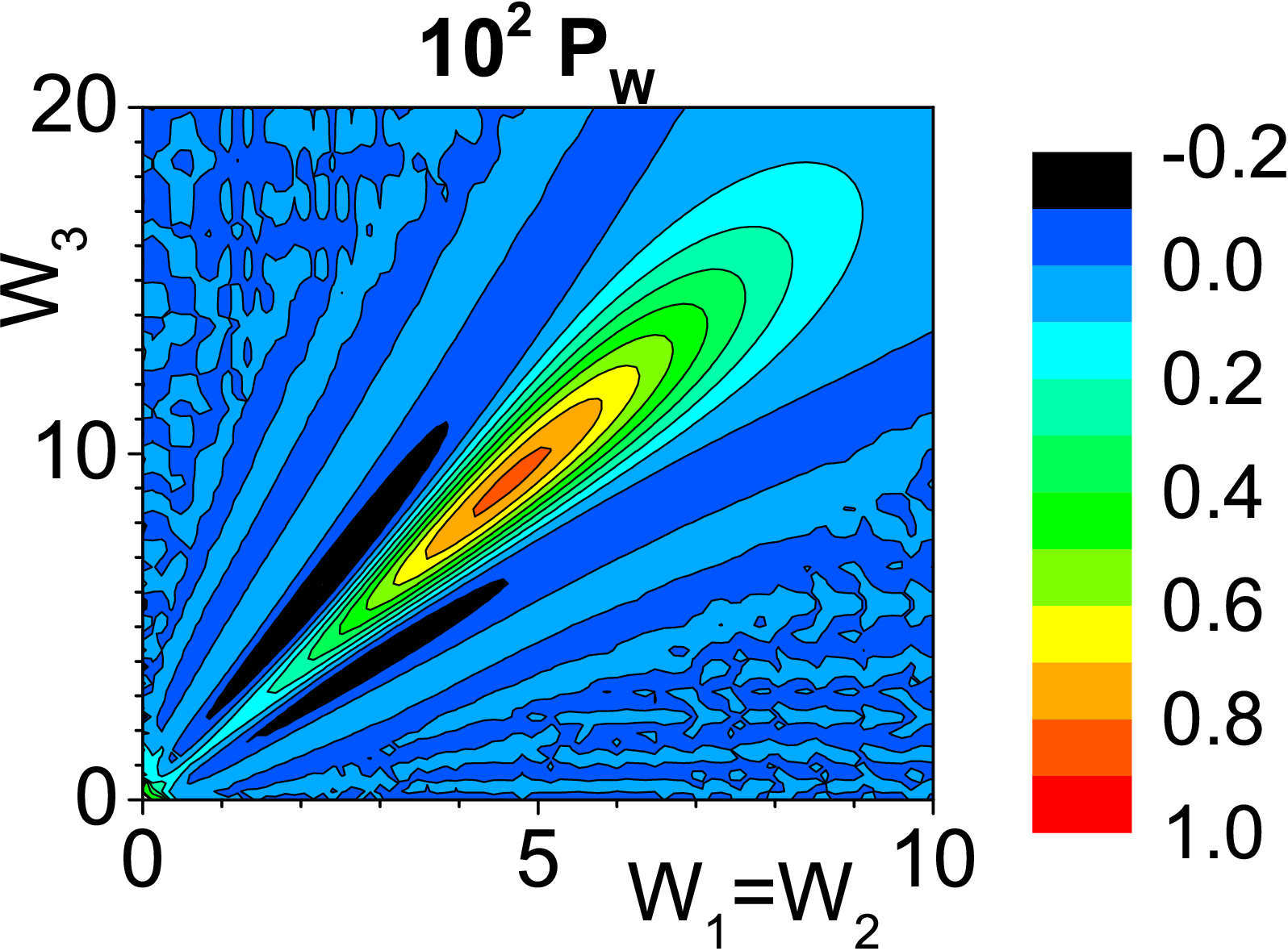}}  \hspace{1mm}
 \resizebox{.47\hsize}{!}{\includegraphics{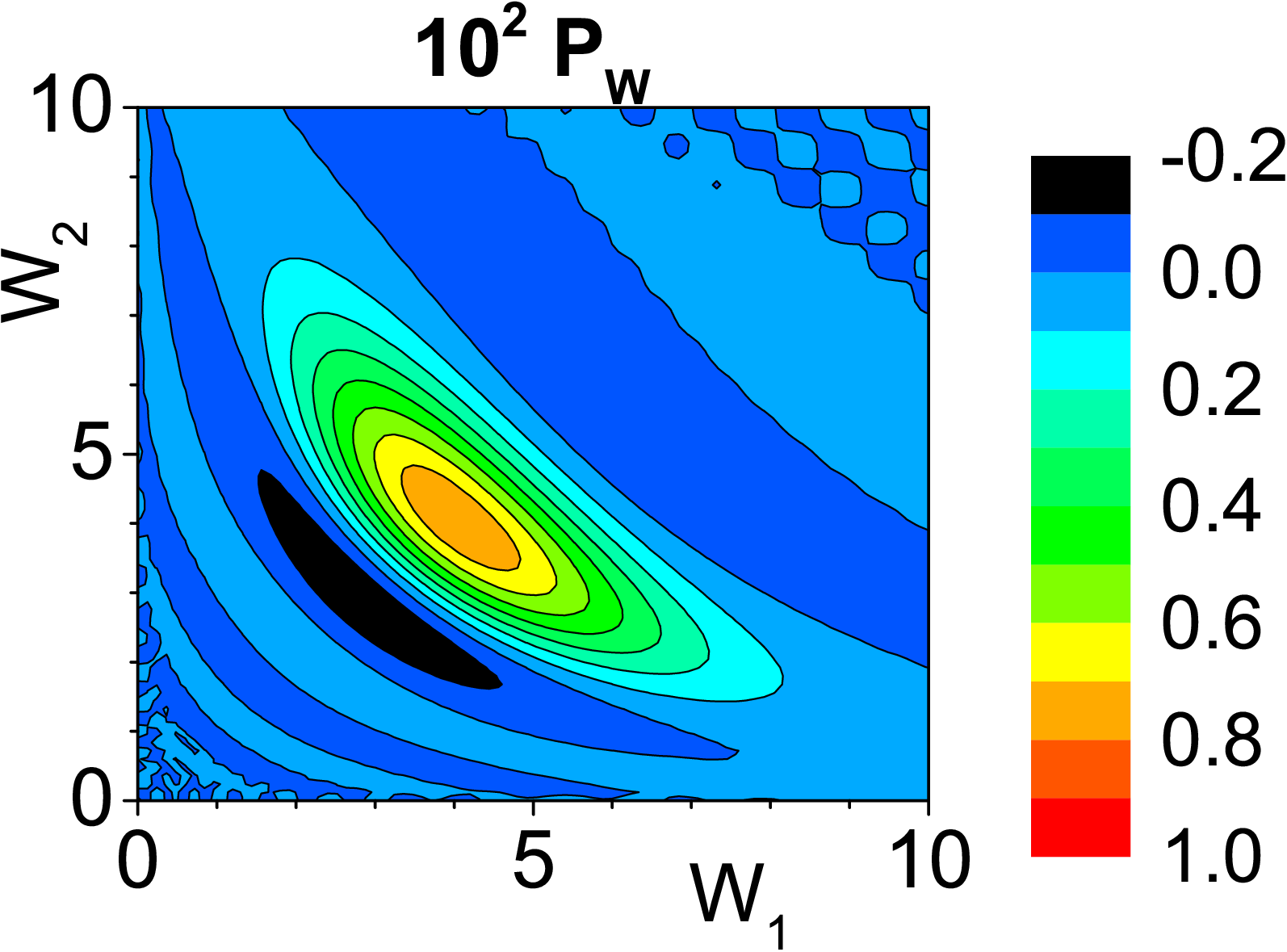}} \\
 \centerline{\small (a) \hspace{.4\hsize} (b)}
 \caption{Quasi-distribution $ P_W(W_1,W_2,W_3) $ of integrated intensities for the
 ordering parameter $ s=0.02 $ and the field obtained by ML approach
 in 2D cuts: (a) $ P_W(W_1=W_2,W_3) $ and (b) $ P_W(W_1,W_2,W_3=7.9) $.}
\label{fig3}
\end{figure}

By definition, the non-classicality of an optical field means that its
quasi-distributions $ P_W $ of integrated intensities attain negative values
for ordering parameters $ s > s_{\rm th} $ where $ s_{\rm th} $ denotes a
threshold value of the ordering parameter. An $ s $-ordered quasi-distribution
$ P_{W,s}({\bf W}) $ of integrated intensities for one effective mode in each
field is obtained by the following formula \cite{Perina1991}:
\begin{eqnarray} 
 & P_{W,s}({\bf W}) = \frac{2^N}{(1-s)^N} \exp\left(-\frac{2\Sigma {\bf W}}{1-s}\right)
   \sum_{{\bf n}={\bf 0}}^{\infty}  \frac{p({\bf n})}{{\bf n}!} & \nonumber \\
 & \times
  \left(\frac{s+1}{s-1}\right)^{\Sigma {\bf n}}  L_{n_1}\left(\frac{4W_1}{1-s^2}\right)
  \ldots L_{n_N}\left(\frac{4W_N}{1-s^2}\right) ; &
\label{33}
\end{eqnarray}
$ L_k $ stand for the Laguerre polynomials \cite{Morse1953}. We
demonstrate the nonclassical behavior of quasi-distribution $ P_W
$ in Fig.~\ref{fig3} where we plot two characteristic cuts for $
s=0.02 $. Whereas the quasi-distribution $ P_W $ creates
hyperbolic structures in planes $ (W_1,W_2) $ (for fixed values of
intensity $ W_3 $), it forms the structure of rays coming from the
point $ (W_1,W_2,W_3) = (0,0,0) $ in plane $ (W_1=W_2,W_3) $ that
is well known for twin beams \cite{PerinaJr2017a}.

Now we apply the appropriate NCCa of Sec.~II for the introduced $ N=3 $
dimensional optical field. We address in turn intensity NCCa, probability NCCa
and hybrid NCCa.

\subsection{Application of intensity non-classicality criteria}

The intensity NCCa containing the lower-order intensity moments
are in general stable when applied to the experimental data. This
is a consequence of the fact that all experimental data are
exploited when the intensity moments are determined. This
contrasts with the probability NCCa where only a rather limited
amount of the experimental data is used for each NCC. It holds in
general for the intensity NCCa that the greater the order of the
moments used in a given NCC is the greater the experimental error
is. Nevertheless, the NCCa containing the lowest-order intensity
moments are usually reliable and very efficient in revealing the
non-classicality.

\begin{figure} 
 \resizebox{.47\hsize}{!}{\includegraphics{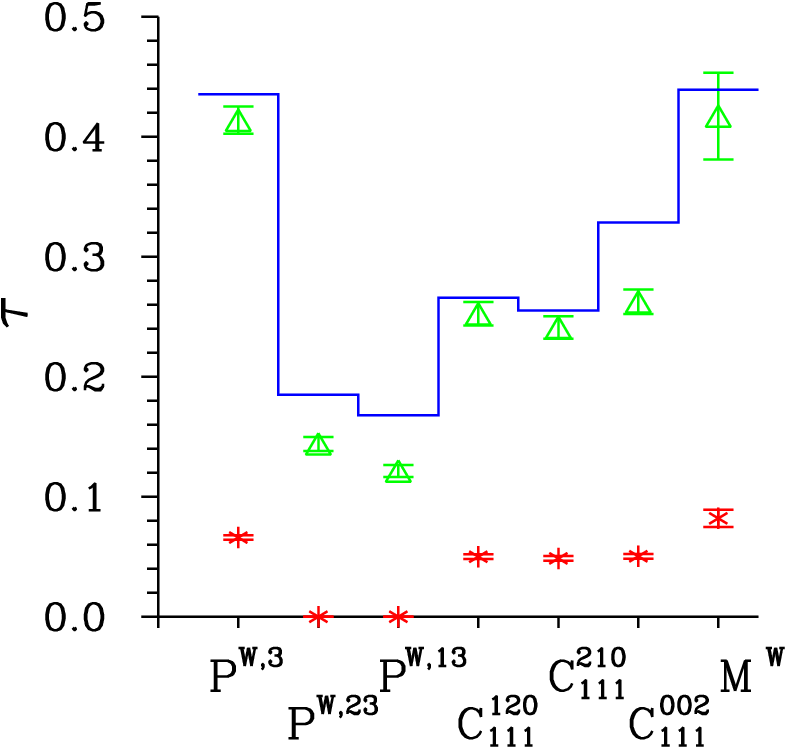}}  \hspace{1mm}
 \resizebox{.47\hsize}{!}{\includegraphics{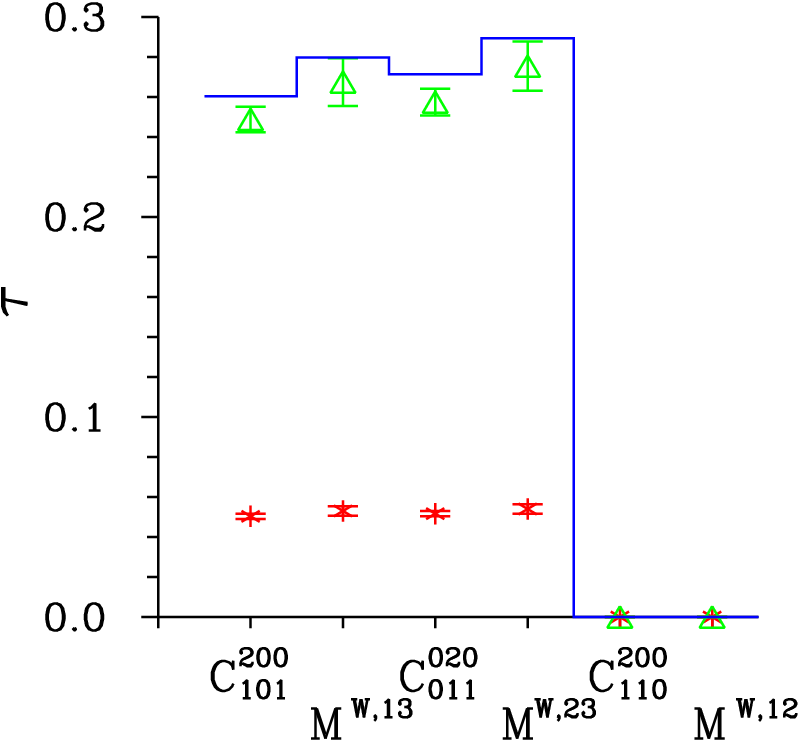}} \\
 \centerline{\small (a) \hspace{.4\hsize} (b)}
 \caption{Non-classicality depths $ \tau $ for (a) 3D and (b) 2D intensity NCCa.
  Isolated symbols with error bars are drawn for the experimental photocount
  histogram (red $ \ast $) and field reconstructed by ML approach (green $ \triangle $); solid blue curves originate in 3D Gaussian
  model.}
\label{fig4}
\end{figure}

The following intensity NCCa (arranged according to the increasing order of
involved intensity moments) derived from the general ones in Eqs.~(\ref{2}),
(\ref{5}) [assuming $ {\bf k} = (1,0,0) $, $ {\bf l} = (0,1,0) $,$ {\bf m} =
(0,0,1) $], (\ref{10}) and (\ref{11}) are capable of identifying the
non-classicality of the analyzed field:
\begin{eqnarray}  
  P^{W,3} &=& \langle (W_1+W_2-W_3)^2\rangle <0, \nonumber \\
  P^{W,13} &=& \langle (-W_1+W_2+W_3)^2 (W_1+W_2-W_3)^2\rangle <0, \nonumber \\
  P^{W,23} &=& \langle (W_1-W_2+W_3)^2 (W_1+W_2-W_3)^2\rangle <0, \nonumber \\
  C_{111}^{120} &=& \langle W_1W_2^2\rangle \langle W_1W_3^2\rangle -
   \langle W_1W_2W_3\rangle^2 <0, \nonumber \\
  C_{111}^{210} &=& \langle W_1^2W_2\rangle \langle W_2W_3^2\rangle -
   \langle W_1W_2W_3\rangle^2 <0, \nonumber \\
  C_{111}^{002} &=& \langle W_1^2W_2^2\rangle \langle W_3^2\rangle -
   \langle W_1W_2W_3\rangle^2 <0, \nonumber \\
  M^W &=& \langle W_1^2\rangle \langle W_2^2\rangle \langle W_3^2\rangle
    + \langle W_1W_2\rangle \langle W_1W_3\rangle \langle W_2W_3\rangle
    \nonumber \\
  & & \mbox{} - \langle W_1^2\rangle \langle W_2W_3\rangle^2 -
    \langle W_2^2\rangle \langle W_1W_3\rangle^2     \nonumber \\
  & & \mbox{} -
    \langle W_3^2\rangle \langle W_1W_2\rangle^2.
\label{34}
\end{eqnarray}
The polynomial NCC $ P^{W,3} $ and the matrix NCC $ M^W $ provide the greatest
values of their NCDs $ \tau $ around 0.4, as it follows from the graph in
Fig.~\ref{fig4}(a).
Whereas the matrix NCCa with their complex moment structures are in general
successful in revealing the non-classicality, the simple polynomial NCC $
P^{W,3} $ complies with the pairwise structure of the analyzed field. The
specific type of pairwise correlations in the analyzed field is also detected
by the fourth-order Cauchy-Schwarz NCCa $ C_{111}^{120} $, $ C_{111}^{210} $
and $ C_{111}^{002} $ though the values of the corresponding NCDs $\tau $ are
smaller. Also the fourth-order polynomial NCCa $ P^{W,13} $ and $ P^{W,23} $
reveal the non-classicality, owing to the involved term $ (W_1+W_2-W_3)^2 $
that they share with the powerful NCC $ P^{W,3} $.

More detailed information about the structure of correlations in the analyzed
field is obtained when the marginal intensity quasi-distributions are analyzed.
The following fourth-order intensity NCCa derived from Eqs.~(\ref{2}) and
(\ref{5}) [assuming $ {\bf k} = (0,0) $, $ {\bf l} = (1,0) $, $ {\bf m} = (0,1)
$]
\begin{eqnarray}   
  C_{101}^{200} &=& \langle W_1^2\rangle \langle W_3^2\rangle -
   \langle W_1W_3\rangle^2 <0, \nonumber \\
  C_{110}^{200} &=& \langle W_1^2\rangle \langle W_2^2\rangle -
   \langle W_1W_2\rangle^2 <0, \nonumber \\
  C_{011}^{020} &=& \langle W_2^2\rangle \langle W_3^2\rangle -
   \langle W_2W_3\rangle^2 <0, \nonumber \\
  M^{W,ij} &=& \langle W_i^2\rangle \langle W_j^2\rangle +
   2\langle W_i\rangle \langle W_j\rangle \langle W_i W_j\rangle \nonumber \\
  & & \mbox{} - \langle W_i\rangle^2 \langle W_j^2\rangle - \langle W_i^2\rangle \langle
  W_j\rangle^2 - \langle W_i W_j\rangle^2 ,\nonumber \\
  & & \mbox{}  \hspace{3mm} (i,j) = (1,2), (1,3)
  (2,3),
\label{35}
\end{eqnarray}
were found useful for this task. The corresponding NCDs $ \tau $ are shown in
Fig.~\ref{fig4}(b). They reveal strong correlations in the marginal fields
(1,3) and (2,3) and no correlation in the marginal field (1,2), in agreement
with the structure of the experimentally generated 3D field. Both the matrix
and the Cauchy-Schwarz NCCa lead to the values of NCDs $ \tau $ around 0.25 for
the fields (1,3) and (2,3). The NCDs $ \tau $ shown in Fig.~\ref{fig4} indicate
slightly greater non-classicality in the field (2,3) compared to the field
(1,3).

The values of NCDs $ \tau $ in Fig.~\ref{fig4} determined for the model
Gaussian field (solid blue curves) are slightly greater than those obtained for
the analyzed field reconstructed by the ML approach (green $ \triangle $). This
reflects the fact that the Gaussian model partly conceals the noise present in
the experimental data. For comparison, we plot in Fig.~\ref{fig4} also the
values of the NCDs $ \tau $ determined directly from the experimental
photocount histogram $ f $ (red $ \ast $). Whereas the NCCa $ P^{W,3} $ and $
M^W $ give the greatest values of NCDs $ \tau $ already for the histogram $ f
$, the NCCa $ P^{W,13} $ and $ P^{W,23} $ do not indicate the non-classicality
in the histogram $ f $; they need stronger and less-noisy fields for successful
application.

\subsection{Probability non-classicality criteria}

In general the probability NCCa are more efficient in revealing the
non-classicality \cite{PerinaJr2017c,PerinaJr2020} compared to their intensity
counterparts. The reason is that they test the field non-classicality locally
via the probabilities in the field photon-number distribution. On the other
hand, the determination of probabilities is more prone to experimental errors
compared to the intensity moments whose determination involves all
probabilities.

Local non-classicality can be investigated using suitable
probability NCCa containing only the probabilities from small
regions. We illustrate this approach by defining the probability
NCCa $ C^p $ and $ M^p $ that are in fact specific groups of the
Cauchy-Schwarz and the matrix NCCa from Eqs.~(\ref{14}) and
(\ref{15}):
\begin{eqnarray}   
  C^p_{\bf n} &=& \min_{ {\bf m}, |{\bf m}-{\bf n}|\le 1} \{
   \bar{C}_{\bf n}^{\bf m} \} < 0, \nonumber \\
  M^p_{\bf n} &=& \min_{{\bf k}, {\bf l}, |{\bf k}-{\bf n}|\le 1, |{\bf l}-{\bf n}|\le 1} \{
   \bar{M}_{\bf kln} \} < 0,
\label{36}
\end{eqnarray}
where $ |{\bf m}-{\bf n}|\le 1 $ stands for the simultaneous conditions $
|m_i-n_i|\le 1 $ for $ i=1,2,3 $. We also consider the probability polynomial
NCCa $ E^{p,13} $ derived from the intensity NCCa in Eq.~(\ref{7}) because
these NCCa are efficient in identifying the non-classicality originating in
photon pairing \cite{PerinaJr2020a}:
\begin{eqnarray}   
  E^{p,13}_{\bf n} &=& \frac{n_1+2}{n_3+1} p(n_1+2,n_2,n_3) +  \frac{n_3+2}{n_1+1}
  p(n_1,n_2,n_3+2) \nonumber \\
  & & \mbox{} - 2p(n_1+1,n_2,n_3+1) < 0.
\label{37}
\end{eqnarray}

\begin{figure} 
 \resizebox{.47\hsize}{!}{\includegraphics{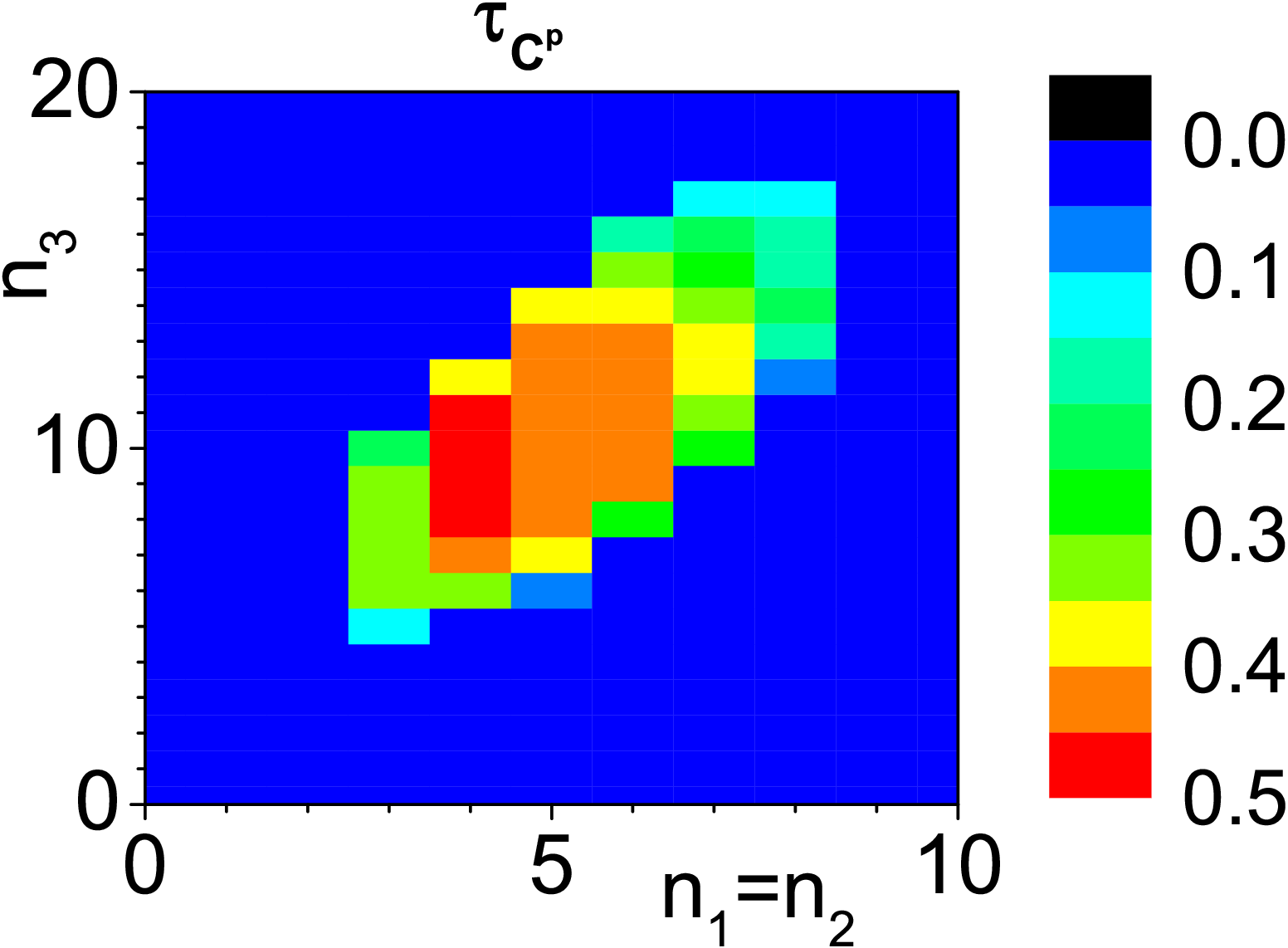}}  \hspace{1mm}
 \resizebox{.47\hsize}{!}{\includegraphics{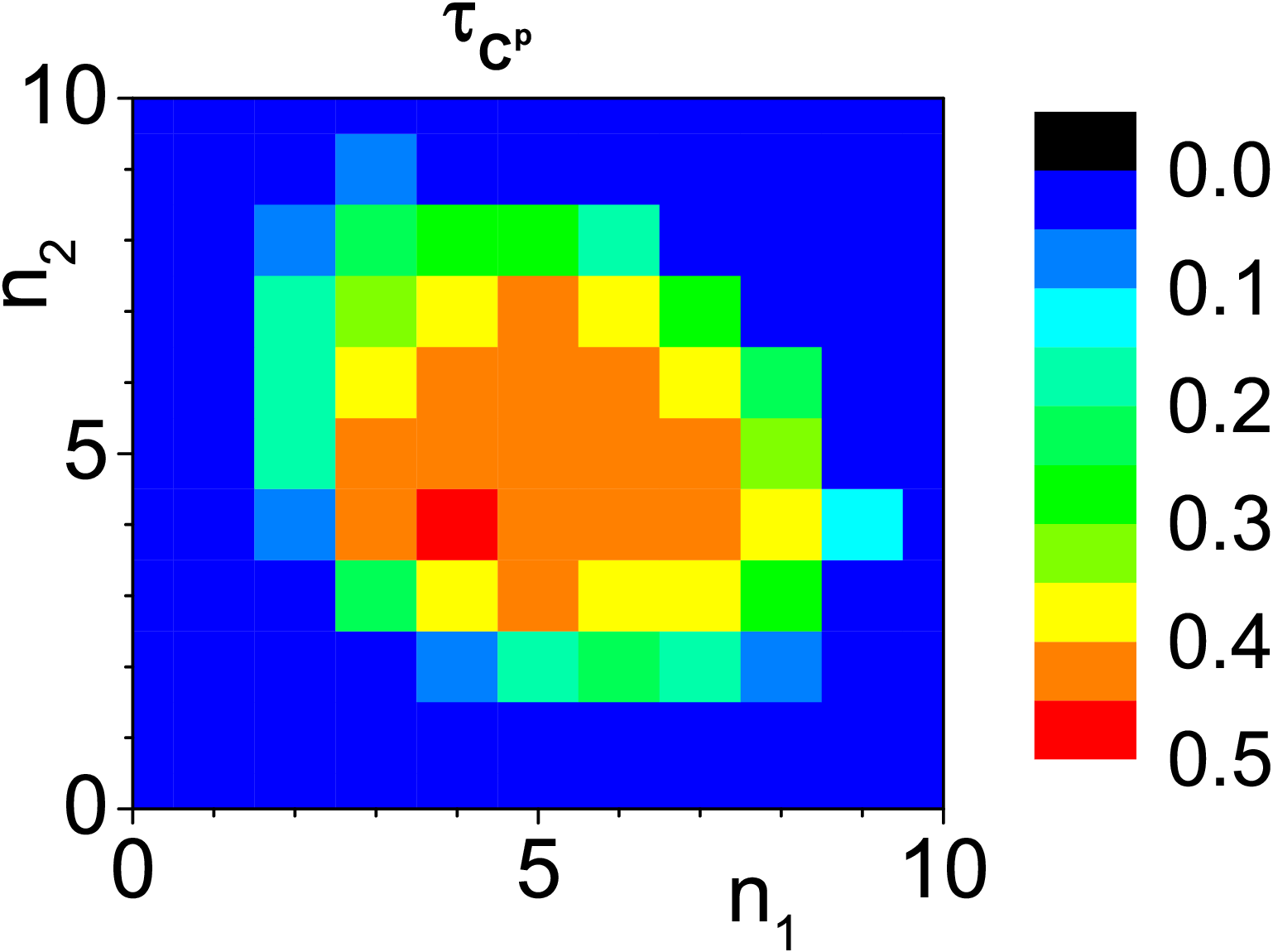}} \\
 \centerline{\small (a) \hspace{.4\hsize} (b)}
 \vspace{2mm}
 \resizebox{.47\hsize}{!}{\includegraphics{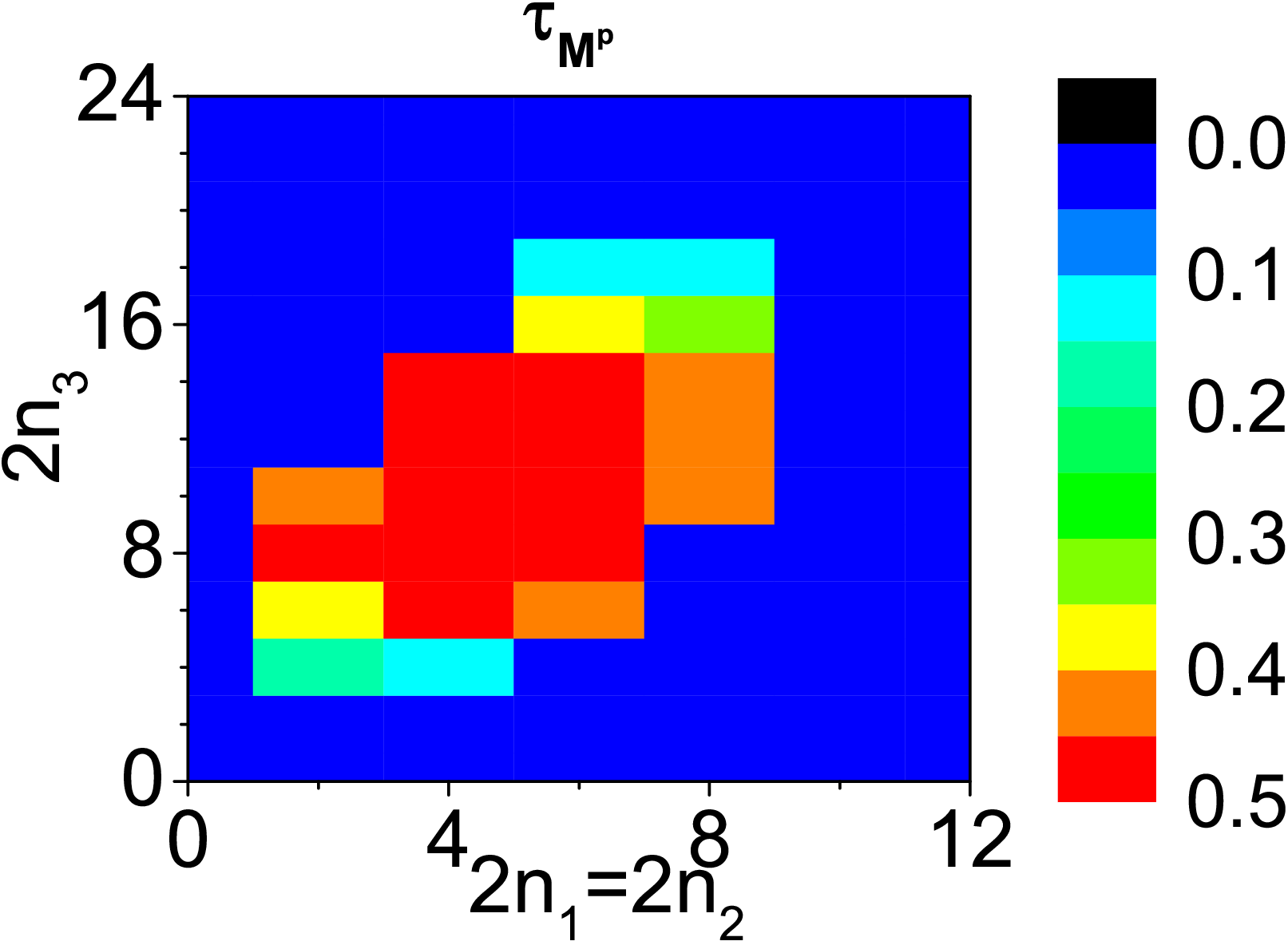}}  \hspace{1mm}
 \resizebox{.47\hsize}{!}{\includegraphics{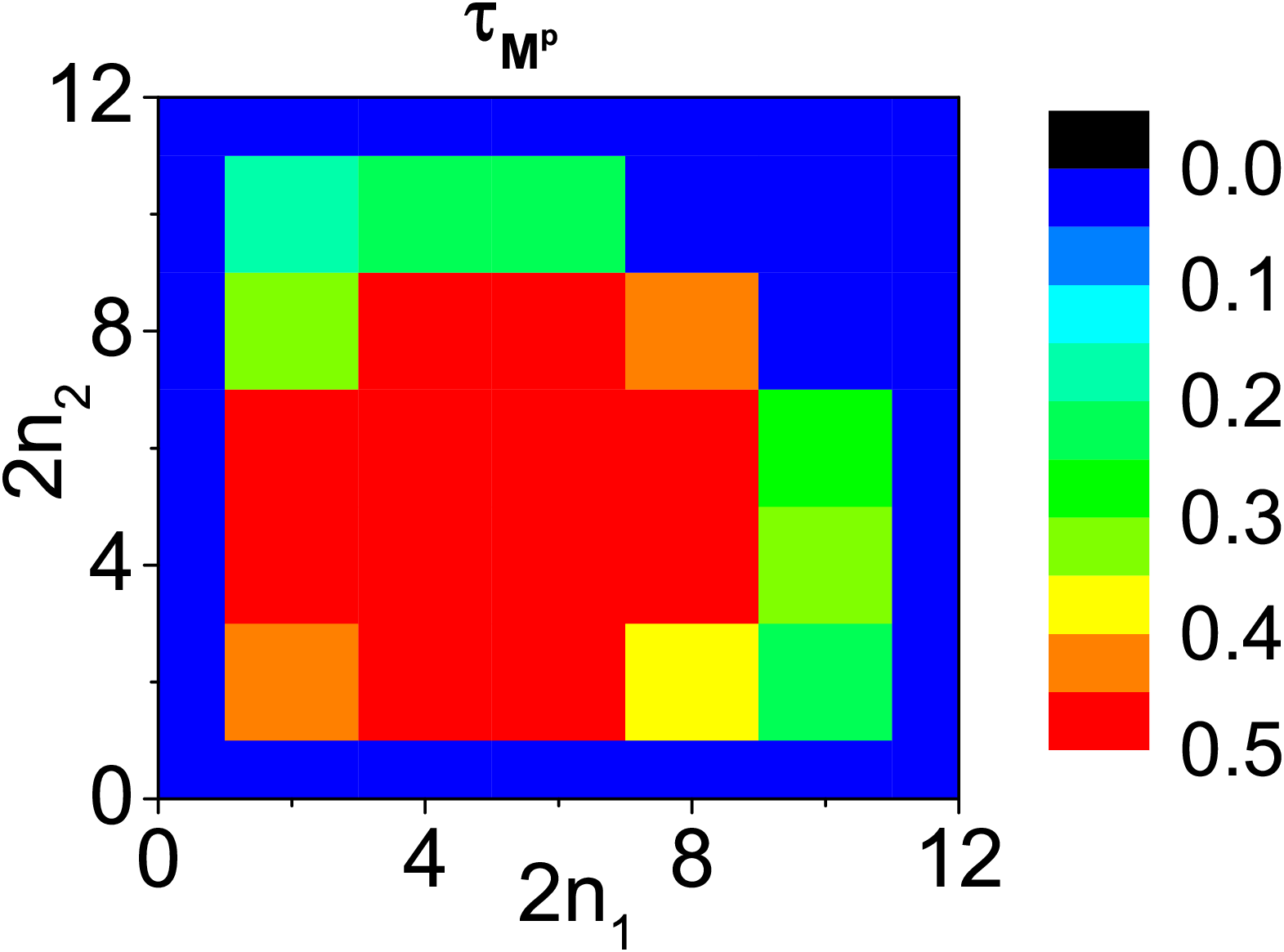}} \\
 \centerline{\small (c) \hspace{.4\hsize} (d)}
 \vspace{2mm}
 \resizebox{.47\hsize}{!}{\includegraphics{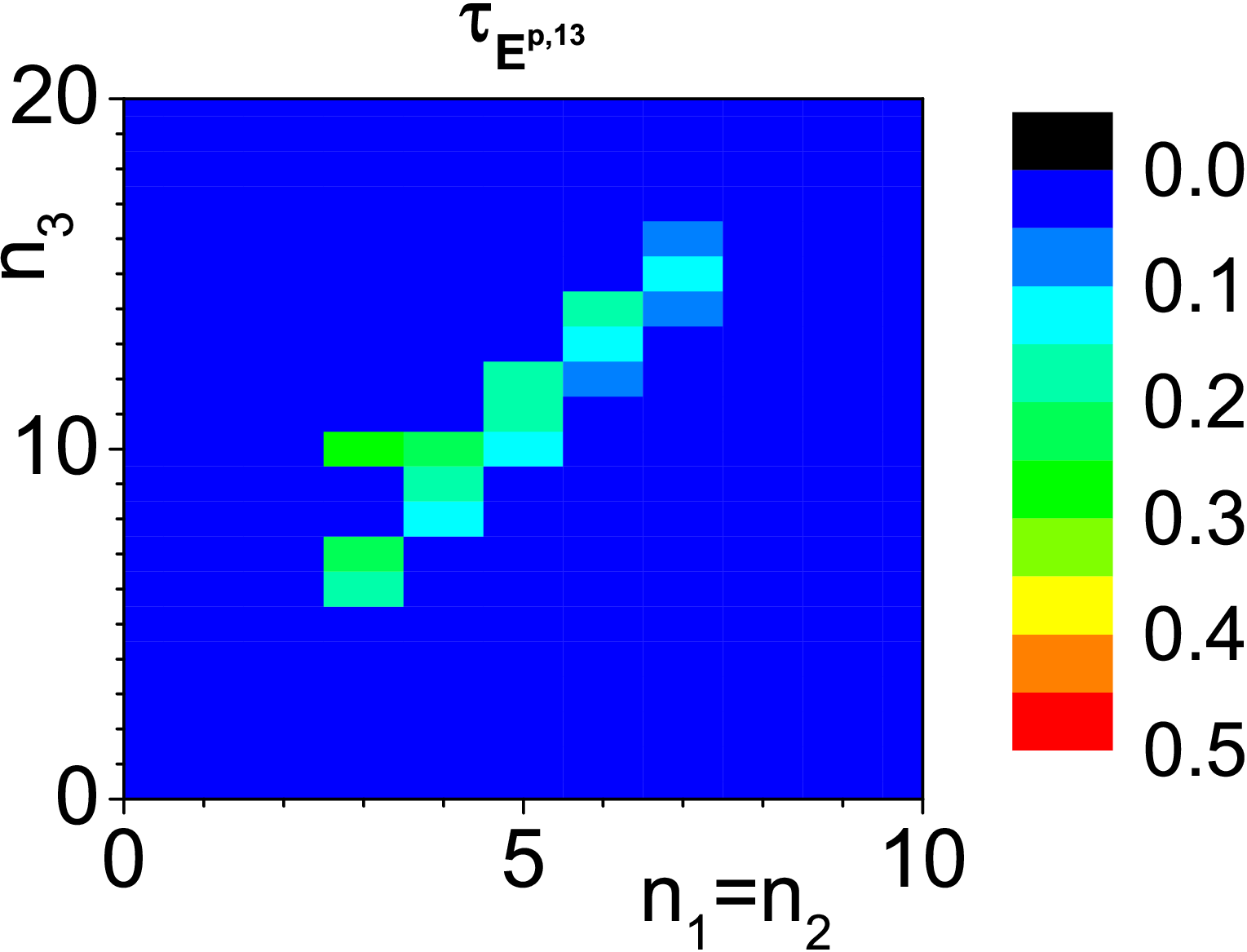}}  \hspace{1mm}
 \resizebox{.47\hsize}{!}{\includegraphics{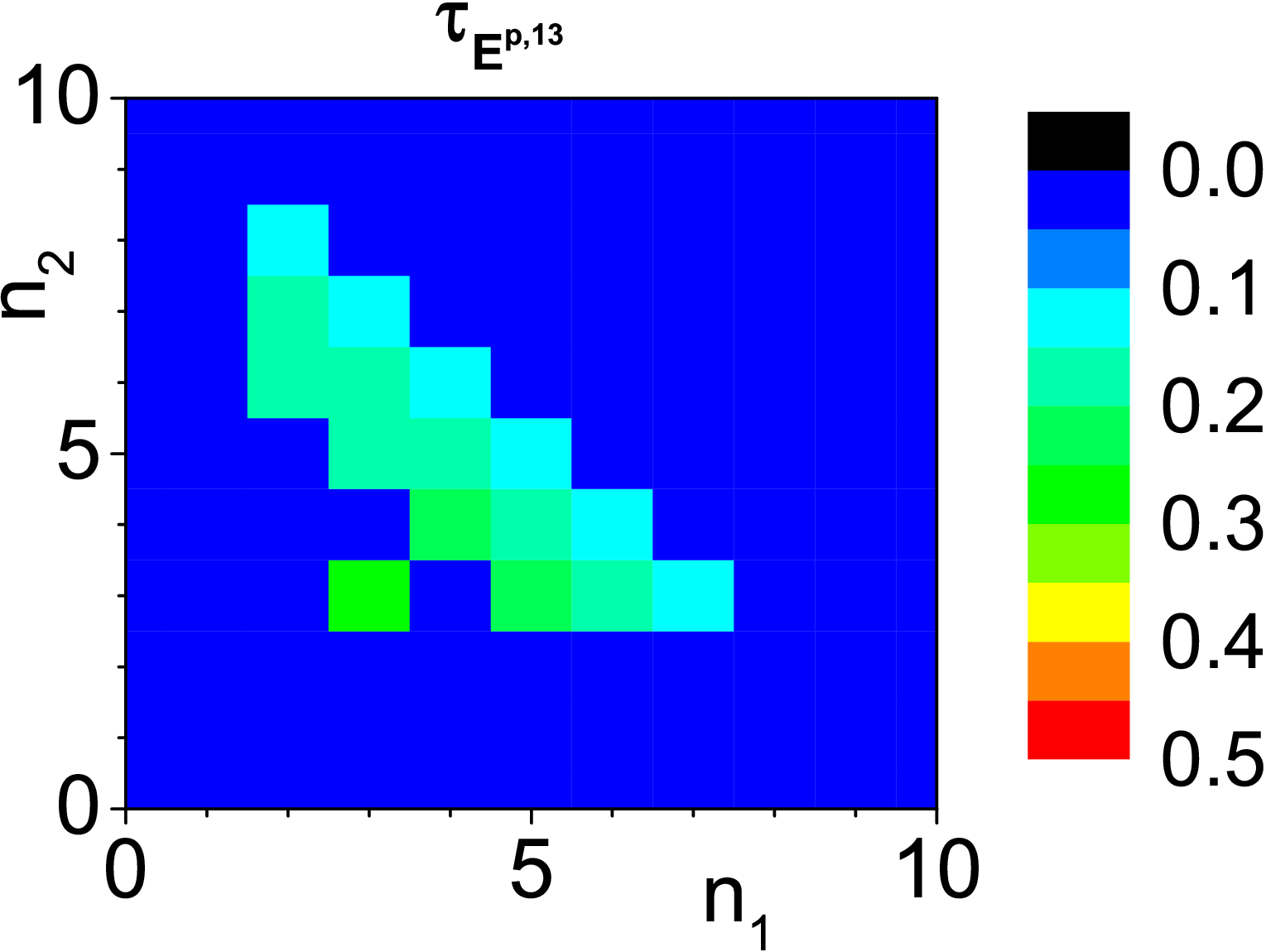}} \\
 \centerline{\small (e) \hspace{.4\hsize} (f)}

 \caption{Non-classicality depths $ \tau $ for probability NCCa (a,b) $ C^p $, (c,d) $ M^p $ and
  (e,f) $ E^{p,13} $ as they depend on photon numbers $ n_1 $, $ n_2 $ and $ n_3 $
  in 2D cuts; $ n_3 = 10 $ in (b,f), $ 2n_3 = 10 $ in (d). Only the
  NCCa for which the mean value of the used probabilities is greater than 0.001 are taken into account.
  Relative errors in (a,b), (c,d) and (e,f) are in turn better than 3~\%, 8~\% and 2~\%.}
\label{fig5}
\end{figure}

The cuts of the probability NCCa $ C^p $, $ M^p $ and $ E^{p,13} $ plotted in
Fig.~\ref{fig5}, that correspond to the cuts of the photon-number distribution
$ p $ in Fig.~\ref{fig2}, reveal that the greatest values of the NCDs $ \tau $
occur in the central part of the 3D photon-number distribution $ p $. These
values drop down as we move towards the photon-number distribution tails.
As documented in the graphs of Fig.~\ref{fig5} the matrix NCCa $
M^p $ perform the best followed by the Cauchy-Schwarz NCCa $ C^p
$; both types of NCCa indicate the greatest values of NCDs $\tau $
close to 0.5. The polynomial NCCa $ E^{p,13} $ give the maximal
values of NCDs $ \tau $ only around 0.3, which is a consequence of
their structure that is sensitive only to photon pairs in the
field 13. Comparing the probability NCCa with their intensity
counterparts, the NCDs $\tau $ are greater by around 0.1 (0.2) for
the matrix (Cauchy-Schwarz) NCCa.

Contrary to the probability NCCa discussed above and containing finite numbers
of probabilities, the Hillery criteria $ H_1 $ and $ H_2 $ from Eqs.~(\ref{24})
and (\ref{27}) contain infinite numbers of probabilities. However, they did not
perform well when analyzing the 3D field reconstructed by the ML approach: Only
the NCC $ H_2 $ applied in 3D provided non-zero value of NCD $ \tau = 0.020\pm
0.001 $. On the other hand, for the model Gaussian field the NCCa $ H_2 $
revealed the non-classicality of 3D field [$ \tau = 0.211 $] as well as the
marginal 2D fields (1,3) [$ \tau = 0.015 $] and (2,3) [$ \tau = 0.050 $]. This
indicates that the Hillery criteria are not suitable for identifying the
non-classicality in experimental photon-number distributions because of the
inevitable noise.

\subsection{Hybrid non-classicality criteria}

The hybrid criteria that contain both probabilities and intensity
moments represent in certain sense a bridge between the intensity
and probability NCCa. Their expected performance in revealing the
non-classicality and resistance against the experimental errors
lie in the middle. On one side they are less efficient but more
stable than the probability NCCa, on the other side they are more
powerful but less stable than the intensity NCCa. With their help,
we can monitor specific aspects of the non-classicality of the
analyzed fields including the experimental issues.

To demonstrate their properties, we first consider the following Cauchy-Schwarz
and the matrix NCCa derived from Eq.~(\ref{21}) [assuming $ {\bf m}_W = (0,2)
$, $ {\bf n}_W = (1,1) $, $ {\bf m}_p = (n_k-m_k) $, $ {\bf n}_p = (n_k) $] and
Eq.~(\ref{5}) converted partly into the probability NCCa via the mapping in
Eq.~(\ref{19}) [assuming $ {\bf k} = (0,0,n_k) $, $ {\bf l} = (1,0,n_k) $, $
{\bf m} = (0,1,n_k) $]:
\begin{eqnarray}  
  C^{pW,k}_{n_k,m_k} &=& \frac{(n_k+m_k)! (n_k-m_k)!}{(n_k!)^2} \langle
   W_i^2\rangle_{p_k(n_k+m_k)} \nonumber \\
  & & \mbox{} \times \langle W_j^2\rangle_{p_k(n_k-m_k)} -
   \langle W_iW_j\rangle_{p_k(n_k)}^2 < 0, \nonumber \\
  M^{pW,k}_{n_k} &=& \langle 1\rangle_{p_k(2n_k)} \langle W_i^2\rangle_{p_k(2n_k)} \langle W_j^2\rangle_{p_k(2n_k)}
    \nonumber \\
  & & \hspace{-10mm} \mbox{} + 2\langle W_i\rangle_{p_k(2n_k)} \langle W_j\rangle_{p_k(2n_k)} \langle W_i W_j\rangle_{p_k(2n_k)} \nonumber \\
  & & \hspace{-10mm} \mbox{} - \langle W_i\rangle^2_{p_k(2n_k)}
   \langle W_j^2\rangle_{p_k(2n_k)}  - \langle W_i^2\rangle_{p_k(2n_k)} \langle
   W_j\rangle^2_{p_k(2n_k)} \nonumber \\
  & & \hspace{-10mm} \mbox{} -  \langle 1\rangle_{p_k(2n_k)} \langle W_i W_j\rangle_{p_k(2n_k)}^2 < 0, \nonumber \\
  & & (i,j,k) = (1,2,3), (2,3,1), (3,1,2).
\label{38}
\end{eqnarray}

\begin{figure} 
 \resizebox{.47\hsize}{!}{\includegraphics{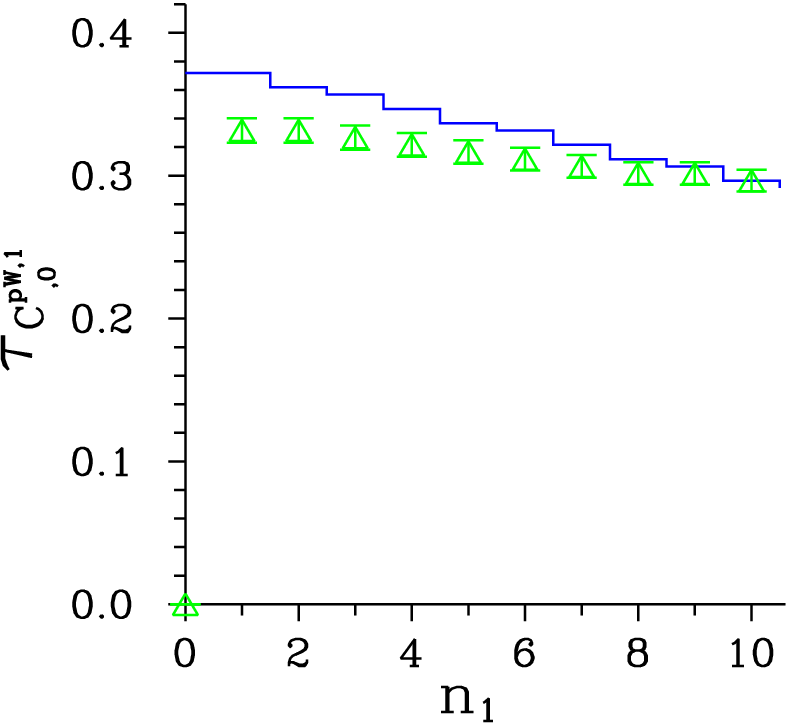}}  \hspace{1mm}
 \resizebox{.47\hsize}{!}{\includegraphics{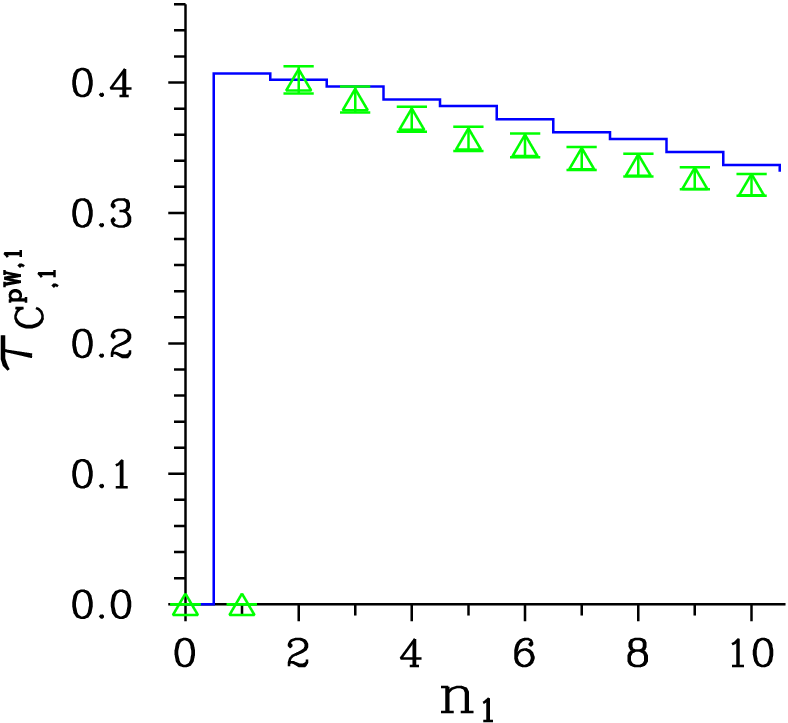}} \\
 \centerline{\small (a) \hspace{.4\hsize} (b)}
 \vspace{2mm}
 \resizebox{.47\hsize}{!}{\includegraphics{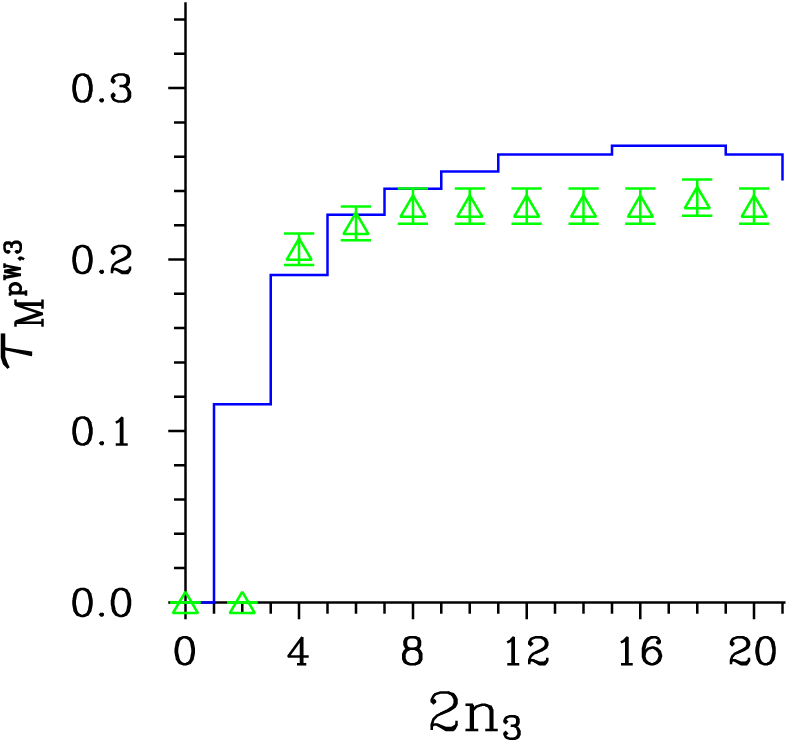}}  \hspace{1mm}
 \resizebox{.47\hsize}{!}{\includegraphics{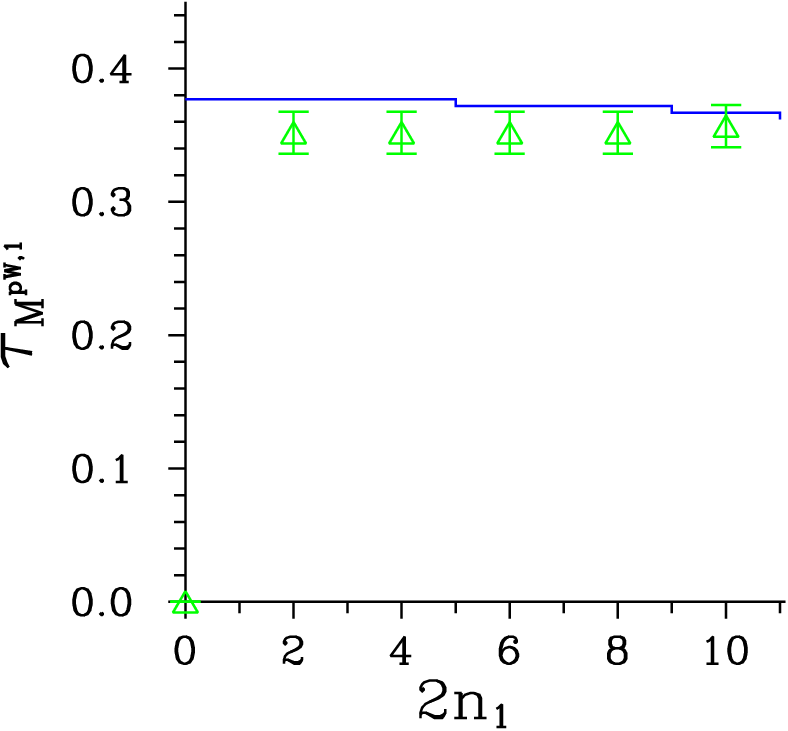}} \\
 \centerline{\small (c) \hspace{.4\hsize} (d)}

 \caption{Non-classicality depths $ \tau $ for hybrid NCCa (a) $ C^{pW,1}_{,0}$,
  (b) $ C^{pW,1}_{,1} $, (c) $ M^{pW,3} $, and (c) $ M^{pW,1} $ as they depend
  on the corresponding photon numbers $ n_1 $, $ 2n_1 $ and $ 2n_3 $. Isolated symbols with error bars
  are plotted for the field reconstructed by ML approach (green $ \triangle $); solid blue curves originate in 3D Gaussian
  model.}
\label{fig6}
\end{figure}

The Cauchy-Schwarz NCC $ C^{pW,k} $ for $ m_k=0 $ represents a 2D intensity NCC
for the field  ($ i,j $) conditioned by the detection of $ n_k $ photons in
field $ k $. The values of NCDs $ \tau $ for the field (2,3) lie around 0.35,
as documented in Fig.~\ref{fig6}(a).
They are greater by around 0.1 compared to the value $ \tau = 0.257\pm 0.007 $
for the NCC $ C^{020}_{011} $ of the marginal field (2,3). This is explained as
follows. The conditional fields occur in the decomposition of the marginal
field (2,3) (with appropriate weights). Composing the marginal field from its
conditional constituents we partly conceal the non-classicality. The
Cauchy-Schwarz NCC $ C^{pW,k} $ in its general form ($ m_k \ne 0 $) involves
the moments of three 2D fields ($ i,j $) conditioned by the detection of $ n_k
-m_k $, $ n_k $ and $ n_k+m_k $ photons in field $ k $. In its general form it
allows to reach even greater values of the NCDs $ \tau $, as demonstrated in
Fig.~\ref{fig6}(b).

Two limiting cases of the behavior of the non-classicality when
composing the field from its conditional constituents are shown in
Figs.~\ref{fig6}(c,d) considering the matrix NCCa $ M^{pW,1} $ and
$ M^{pW,3} $ from Eq.~(\ref{36}). Whereas the hybrid NCCa $
M^{pW,3} $ indicate the NCDs $ \tau $ around 0.22 for the
conditional fields (1,2) in Fig.~\ref{fig6}(c), the marginal field
(1,2) is classical ($ M^{W,23} = 0 $). On the other hand, the
hybrid NCCa $ M^{pW,1} $ assign the NCDs $ \tau $ around 0.35 for
the conditional fields (2,3) and similar value $ \tau = 0.28\pm
0.01 $ is obtained for the marginal field (2,3) applying the NCC $
M^{W,23} $. We note that, in Fig.~\ref{fig6}, similarly as in the
case of intensity NCCa, the values of NCDs $\tau $ obtained from
the model Gaussian field are slightly greater than those
characterizing the field reconstructed by the ML approach.

As another example, we consider the following hybrid
Cauchy-Schwarz and matrix criteria $ C^{Wp,i} $ and $ M^{Wp,i} $
obtained from Eq.~(\ref{21}) [assuming $ {\bf m}_W = (0) $, $ {\bf
n}_W = (1) $, $ {\bf m}_p = {\bf n}_p = (n_j,n_k) $] and
Eq.~(\ref{5}) converted into the probability NCCa [assuming $ {\bf
k} = (0,n_j,n_k) $, $ {\bf l} = (1,n_j,n_k) $, $ {\bf m} =
(2,n_j,n_k) $] using the mapping in Eq.~(\ref{19}):
\begin{eqnarray}  
  C^{Wp,i}_{n_j,n_k} &=& \langle W_i^2\rangle_{p_{jk}(n_j,n_k)}  \langle 1\rangle_{p_{jk}(n_j,n_k)}
   \nonumber \\
  & & \mbox{} - \langle W_i\rangle_{p_{jk}(n_j,n_k)}^2 < 0, \nonumber \\
  M^{Wp,i}_{n_j,n_k} &=& \langle 1\rangle_{p_{jk}(2n_j,2n_k)}
   \langle W_i^2\rangle_{p_{jk}(2n_j,2n_k)} \langle W_i^4\rangle_{p_{jk}(2n_j,2n_k)}\nonumber \\
  & & \hspace{-10mm} \mbox{} + 2\langle W_i\rangle_{p_{jk}(2n_j,2n_k)} \langle W_i^2\rangle_{p_{jk}(2n_j,2n_k)}
   \langle W_i^3\rangle_{p_{jk}(2n_j,2n_k)} \nonumber \\
  & & \hspace{-10mm} \mbox{} - \langle 1\rangle_{p_{jk}(2n_j,2n_k)}
   \langle W_i^3\rangle_{p_{jk}(2n_j,2n_k)}^2 - \langle W_i\rangle_{p_{jk}(2n_j,2n_k)}^2
   \nonumber \\
  & & \hspace{-10mm} \mbox{} \times \langle W_i^4\rangle_{p_{jk}(2n_j,2n_k)} - \langle
   W_i^2\rangle_{p_{jk}(2n_j,2n_k)}^3 < 0, \nonumber \\
  & & (i,j,k) = (1,2,3), (2,3,1), (3,1,2).
 \label{39}
\end{eqnarray}
The hybrid NCCa in Eq.~(\ref{39}) represent 1D intensity NCCa for the field $ i
$ conditioned by simultaneous detection of $ n_j $ photons in field $ j $ and $
n_k $ photons in field $ k $.

\begin{figure}[t] 
 \resizebox{.47\hsize}{!}{\includegraphics{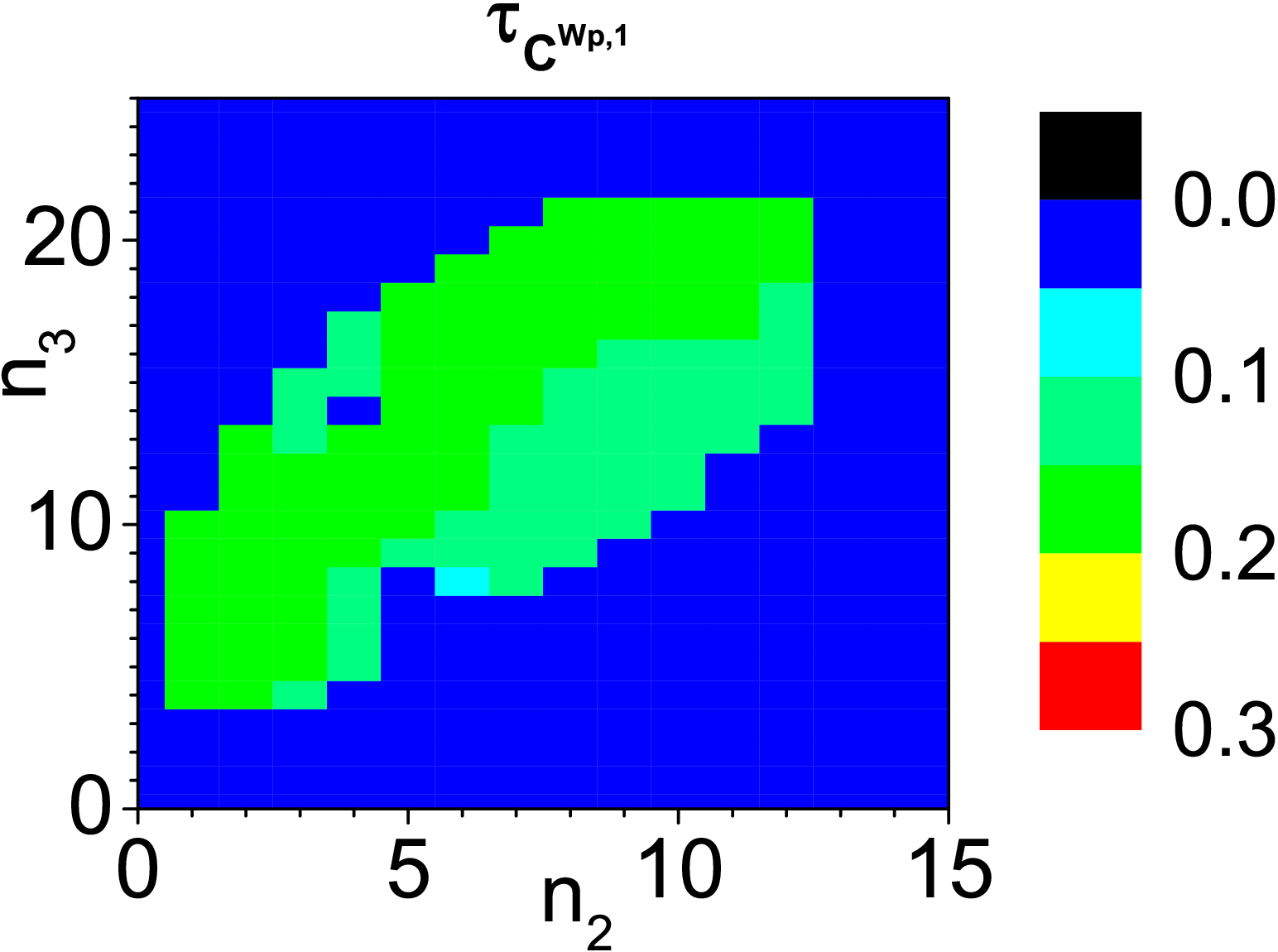}}  \hspace{1mm}
 \resizebox{.47\hsize}{!}{\includegraphics{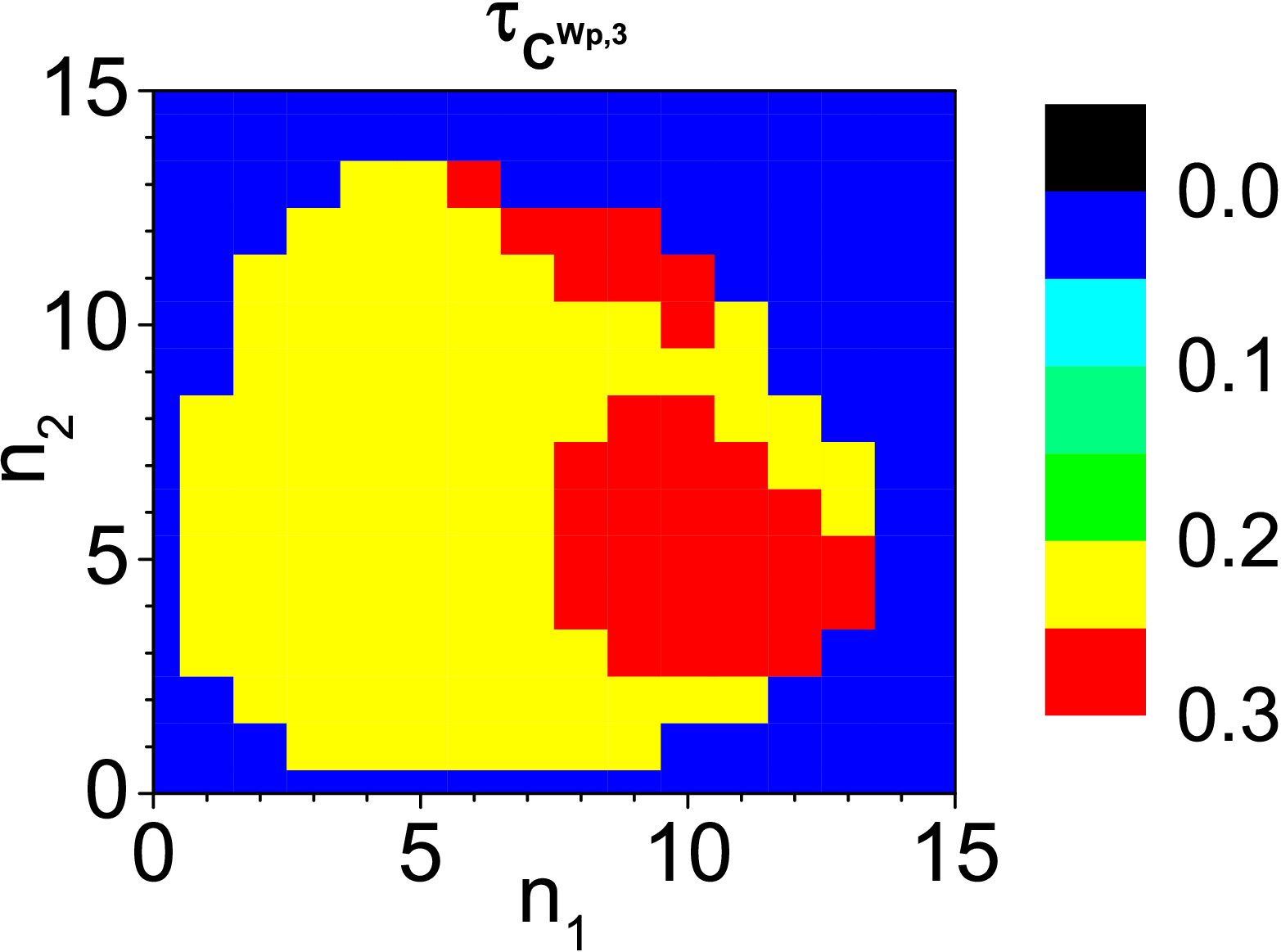}} \\
 \centerline{\small (a) \hspace{.4\hsize} (b)}
 \vspace{2mm}
 \resizebox{.47\hsize}{!}{\includegraphics{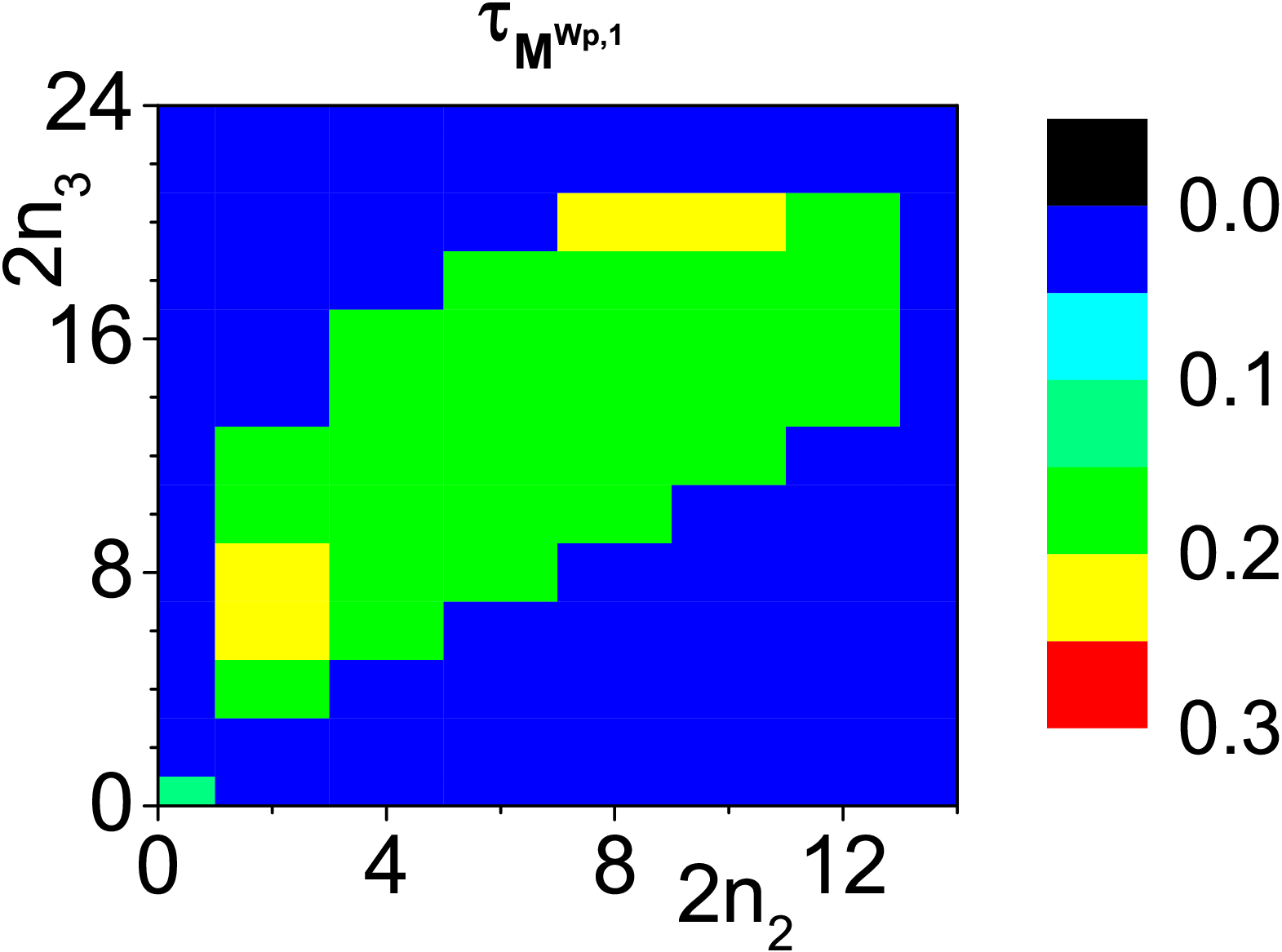}}  \hspace{1mm}
 \resizebox{.47\hsize}{!}{\includegraphics{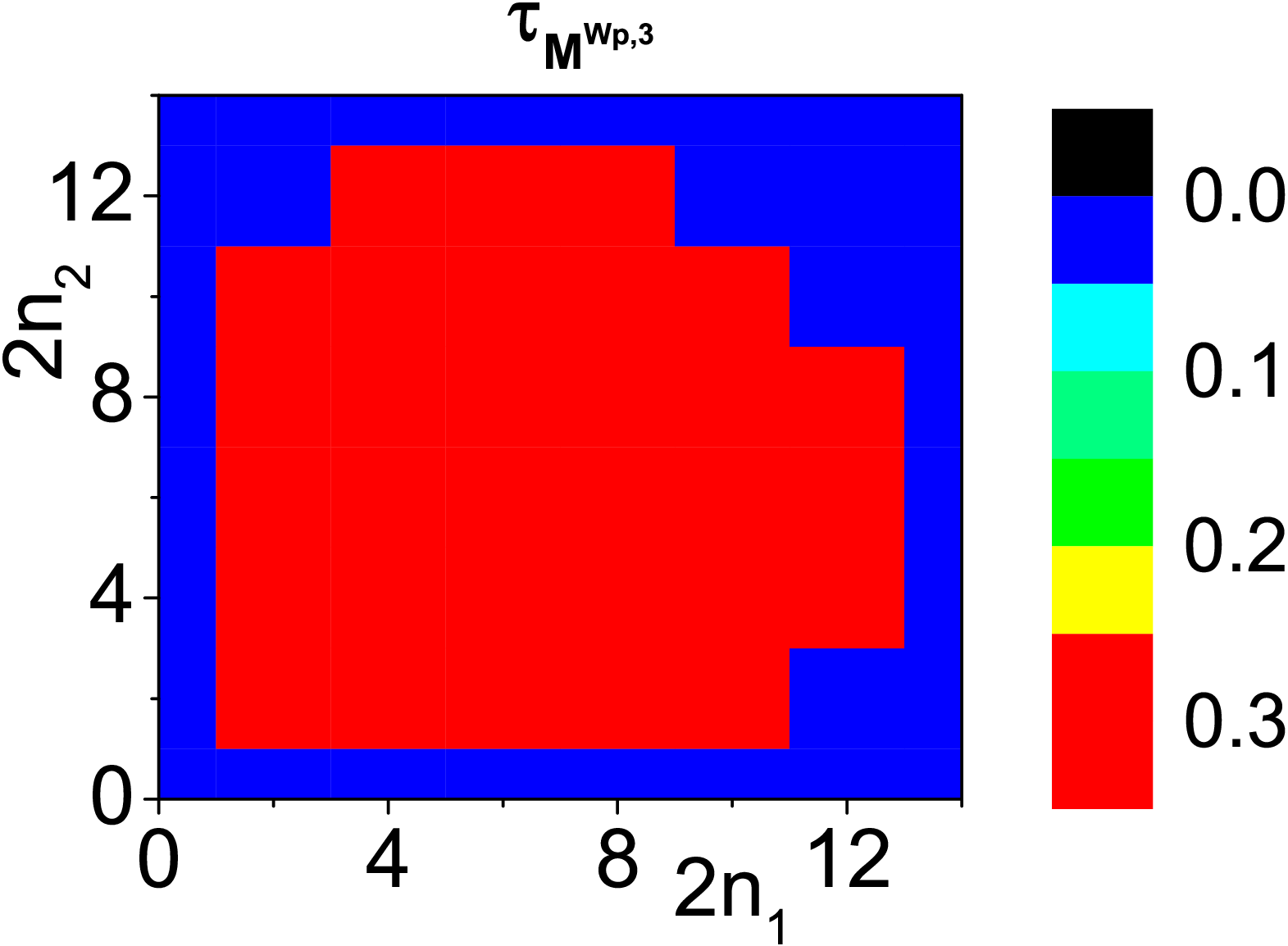}} \\
 \centerline{\small (c) \hspace{.4\hsize} (d)}

 \caption{Non-classicality depths $ \tau $ for hybrid NCCa (a) $ C^{Wp,1}$,
  (b) $ C^{Wp,3} $, (c) $ M^{Wp,1} $, and (c) $ M^{Wp,3} $ as they depend
  on the corresponding photon numbers $ n_1 $, $ n_2 $, $ n_3 $, $ 2n_1 $, $ 2n_2 $, and $ 2n_3 $.
  Relative errors in (a,b) [(c,d)] are better than 3~\% [15~\%].}
\label{fig7}
\end{figure}

For the analyzed 3D field, the conditional fields 1 are less
nonclassical than the conditional fields 3, as evidenced in the
graphs of Fig.~\ref{fig7} showing the NCDs $ \tau $ of the NCCa $
C^{Wp,1} $ and $ M^{Wp,1} $ belonging to conditional fields 1 and
$ C^{Wp,3} $ and $ M^{Wp,3} $ quantifying the non-classicality of
conditional fields 3. The NCCa $ C^{Wp,1} $ and $ M^{Wp,1} $ ($
C^{Wp,3} $ and $ M^{Wp,3} $) assign the values of NCDs $ \tau $
around 0.2 (0.3) for the conditional fields 1 (3).
This asymmetry originates in the structure of the analyzed 3D
field. Whereas the non-classicality in conditional fields 1 is
caused by photon pairs residing in the fields 13 and 23 whose
numbers are chosen by 'two-step' post-selection based on the
detection of given numbers of photons in the fields 2 and 3, the
non-classicality of conditional fields 3 has its origin in both
types of photon pairs (residing in fields 13 and 23) and
independent post-selections requiring the detection of given
numbers of photons in the fields 1 and 2.

\section{Conclusions}

Using the Cauchy-Schwarz inequality, nonnegative quadratic forms,
the majorization theory and nonnegative polynomials we have
formulated large groups of non-classicality criteria for general $
N $-dimensional optical fields. The non-classicality criteria were
written in intensity moments, probabilities of photon-number
distributions and a specific hybrid form that simultaneously
includes both intensity moments and probabilities. The derived
non-classicality criteria were decomposed into the simplest
building blocks and then mutually compared. The fundamental
non-classicality criteria suitable for application were
identified. As a special example, an $ N $-dimensional form of the
Hillery non-classicality criteria was derived.

Discussing the transformation of intensity moments and
photon-number distributions between different field-operator
orderings, quantification of the non-classicality based on these
criteria and using the non-classicality depth was accomplished.

The properties as well as the performance of the derived
non-classicality criteria were demonstrated considering an
experimental 3-dimensional optical field containing two types of
photon pairs. It was shown that the intensity non-classicality
criteria are both efficient in revealing the non-classicality and
robust with respect to experimental errors. The ability of the
probability non-classicality criteria to provide insight into the
distribution of non-classicality across the profile of
photon-number distribution was demonstrated. The hybrid
non-classicality criteria were presented as a useful alternative
to the intensity and probability non-classicality criteria.

The analyzed experimental example proved that the non-classicality
criteria represent a very powerful tool in identifying and
quantifying the non-classicality in its various forms. The derived
non-classicality criteria are versatile and as such they can be
successfully applied to any optical field.

\acknowledgements \noindent J.P. Jr., V.M., R.M. and O.H. thank GA
\v{C}R project No.~18-08874S.


%

\end{document}